\documentstyle[10pt,aaspp4,flushrt]{article}
\def\etal{\it et al. \rm }
\begin{document}

\title{Ages and Metallicities of Fornax Dwarf Ellipticals}

\author{Karl Rakos}
\affil{Institute for Astronomy, University of Vienna, A-1180, Wien, Austria;
karl.rakos@chello.at}

\author{James Schombert}
\affil{Department of Physics, University of Oregon, Eugene, OR 97403;
js@abyss.uoregon.edu}

\author{H.M. Maitzen}
\affil{Institute for Astronomy, University of Vienna, A-1180, Wien, Austria}

\author{Sinisa Prugovecki}
\affil{University of Zagreb, Croatia}

\author{Andrew Odell}
\affil{Department of Physics and Astronomy, Northern Arizona University, Box 6010,
Flagstaff, AZ 86011; andy.odell@nau.edu}

\begin{abstract}

Narrow band photometry is presented on 27 dwarf ellipticals in the Fornax
cluster.  Calibrated with Galactic globular cluster data and
spectrophotometric population models, the colors indicated that dwarf
ellipticals have a mean [Fe/H] of $-1.00\pm0.28$ ranging from $-$1.6 to
$-$0.4.  The mean age of dwarf ellipticals, also determined
photometrically, is estimated at 10$\pm$1 Gyrs compared to 13 Gyrs for
bright Fornax ellipticals.  Comparison of our metallicity color and Mg$_2$
indices demonstrates that the [Mg/Fe] ratio is lower in dwarf ellipticals
than their more massive cousins, which is consistent with a longer
duration of initial star formation to explain their younger ages.  There
is a increase in dwarf metallicity with distance from the Fornax cluster
center where core galaxies are, on average, 0.5 dex more metal-poor than
halo dwarfs. In addition, we find the halo dwarfs are younger in mean age
compared to core dwarfs.  One possible explanation is that the
intracluster medium ram pressure strips the gas from dwarf ellipticals
halting star formation (old age) and stopping enrichment (low metallicity)
as they enter the core. 

\end{abstract}


\section{INTRODUCTION}

Objects composed of single stellar populations (SSP; i.e. globular
clusters) or a composite of SSP's (i.e. ellipticals) present special
circumstances for the study of the evolution of stellar populations.  A
SSP is a stellar population of instantaneous origin at a particular
formation time with a single metallicity.  For many galaxies, the
resolution of the integrated population into a series of composite SSP's
is impossible due to the presence of ongoing star formation.  However, for
systems that have exhausted their gas supply many Gyrs ago, it may be
possible to untangle the underlying population with some simple
assumptions.  The canonical interpretation of the spectroscopy and
photometric data for ellipticals is that they are a composite of a series
of SSP's with a small spread in age that represents the time from the
initial burst of star formation (Charlot \& Bruzual 1991), although there
is new data to indicate that a significant number of field ellipticals
have a large spread in age (see Trager \etal 2000).  Due to the fact that
the initial star formation occurs over some finite duration (i.e. not
simultaneous), enrichment of the ISM by supernovae and stellar winds will
alter the successive generation of stars producing a smooth increase in
mean metallicity.

Only recently have we achieved evolution models, with the proper input
physics and stellar libraries of metal-poor to super-metal-rich stars,
which can follow the photometric changes due to enrichment from previous
populations (e.g. Bruzual \& Charlot 2000).  The two most prominent
observables to the effects of population evolution in ellipticals are the
mass-metallicity effect, the higher mean metallicity for high mass
galaxies, and the existence of metallicity gradients.  The very existence
of metallicity gradients demonstrates that some process of chemical
evolution occurs since a gradient develops by the freezing of the
metallicity as a function of time in the orbits of the first generation
stars (i.e.  oldest and metal-poor) at the outer radii and most enriched
in the core.

Any investigation into the stellar populations in ellipticals must focus
on the separation of the age and metallicity to the underlying stars as
the two key parameters.  Unfortunately, slight changes in age and
metallicity operate in the same direction of spectroevolutionary parameter
space, the well-known age-metallicity degeneracy.  Recently, the
age-metallicity degeneracy has been broken with the use of spectral line
indices (Worthey 1994), but the problem of light element enrichment
plagues the interpretation of line results (Worthey 1998).  While spectral
indices are usually considered superior to colors in the determination of
age and metallicity factors, narrow band colors have a distinct advantage
when the expected age differences will be small and, thus, higher S/N of
the kind offered by integrated color will be required for any spatial
coverage.  In addition, many of the best line indices (H and K Ca, Mg$_2$)
suffer from subtle effects (e.g. $\alpha$ enhancement to non-solar values)
that make interpretation of age and metallicity model dependent.

Most of what we have determined about the stellar populations in
ellipticals has centered on high mass, bright ellipticals while their
lesser cousins, the dwarf ellipticals, have received little attention.
Dwarf ellipticals are, by definition, low in apparent magnitude and the
primary difficulty in obtaining information about their stellar
populations is because their surface brightnesses are very low making
spectral observations difficult regardless of distance.  This is
unfortunate because dwarf galaxies are important to our understanding of
galaxy formation and evolution.  For example, hierarchical models of
structure formation hypothesize that all galaxies are constructed from the
merger of smaller mass dwarfs.  Even with their low surface brightnesses,
there is no shortage in the number of dwarf ellipticals to study since
they make up a significant fraction of the members of rich clusters
(Ferguson 1994).  There have been numerous studies of the properties of
dwarfs (e.g. Bothun \etal 1986); however, there is no solid consensus on
the interpretation of the stellar content of dwarfs.  For example, in an
experiment applied by Arimoto (1996), the spectral templates of dwarf
ellipticals was distributed to numerous population synthesis groups.  The
results concluded that dwarf ellipticals were either 1) old and
metal-rich, 2) old and slightly metal-poor ([Fe/H] $\approx -0.3$) or 3)
young (3-5 Gyrs) and solar metallicity.

For the last 15 years, our group has used a narrow band filter system to
explore the color evolution of galaxies in distant clusters.  This filter
system, a modified Str\"omgren color system, was used initially to follow
the color evolution of ellipticals to a redshift of 0.9 (Rakos \&
Schombert 1995).  However, the Str\"omgren colors were also found to be a
sharp discriminator of starforming galaxies in clusters (the
Butcher-Oemler effect) and, most importantly, several of the indices were
found to have good correlations with metallicity in a study of color
gradients in ellipticals (Schombert \etal 1993).  The color system
displayed promise as an avenue to investigate the properties of SSP's
without the need for extensive model interpretation.

\begin{figure}
\plotfiddle{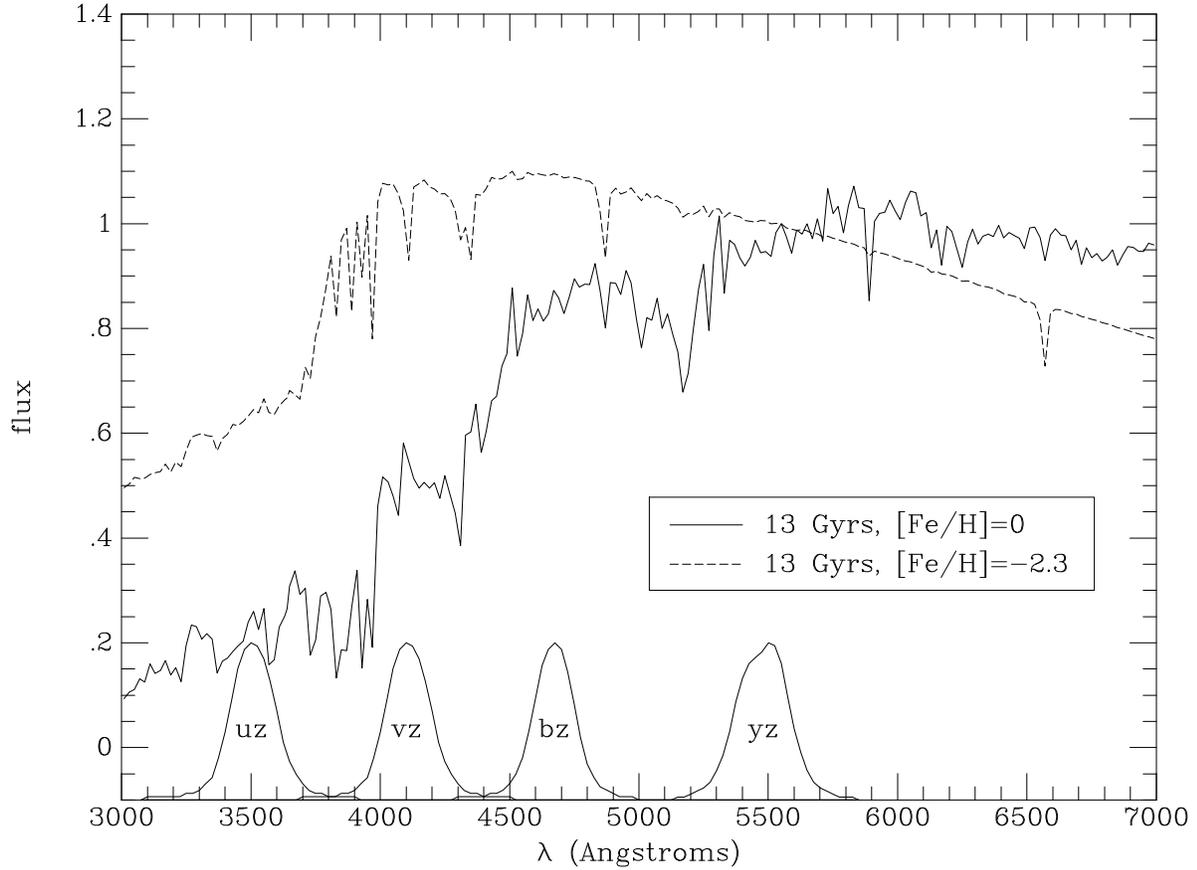}{9.0truecm}{-90}{65}{65}{-270}{360}
\caption{SSP models from BC2000.  The models for a 13 Gyr metal-poor and a
13 Gyr metal-rich population are shown normalized to the 5500\AA\ flux.
The filter response curves for $uz$, $vz$, $bz$ and $yz$ are shown.  The
effects of increased line blanketening and cooler mean temperatures are
evident in the stronger spectral features plus shift with the peak
luminosity to longer wavelengths.}
\end{figure}

Our goal for this paper is twofold; 1) to demonstrate that narrow band
filters are effective in discriminating age and metallicity for single
generation objects and 2) use our narrow band system to compare the
stellar populations in dwarf and giant ellipticals.  While spectral
indices are superior in the detailed information they provide, our goal is
to develop a system that can be used for low surface brightness and/or
distant objects where spectroscopy is impractical.

Due to the need to connect our filter system to the Galactic globular
cluster age and metallicity system, plus demonstrate the effects of
multi-metallicity SSP's on the integrated colors of ellipticals, the
presentation of the data is lengthy.  However, we have divided the paper
into the following topics, comparison to globular clusters, comparison to
SSP models, empirical age \& metallicity calibration and discussion of
the properties of dwarfs compared to bright ellipticals, in order to allow
the reader to focus on the sections of their interest.  Most of our
results are model independent; however, we have made every effort to
single out the observationally secure conclusions from the ones requiring
theoretical assistance.  

\section{OBSERVATIONS}

\subsection{Str\"omgren filter system}

The data for this project was collected using a modified Str\"omgren
filter system.  The original Str\"omgren $uvby$ filter system and its
extensions (Str\"omgren 1966, Wood 1969) were developed to provide the
astronomical community with an accurate means of measuring the
temperature, chemical composition and surface gravity of stars, without
resorting to high resolution spectroscopy.  This is achieved by selecting
center wavelengths to cover regions of the stellar spectral energy
distributions (SED) which are sensitive to the above properties. This
provided an efficient and quantifiable method of sampling stellar types
without the use of qualitative spectral classification (e.g. the MK
scheme).

In the late 1980's, our team developed a modified Str\"omgren system
(called $uz,vz,bz,yz$) which was a rest frame narrow band filter system
used, primarily, to investigate the spectrophotometric evolution of
distant galaxies.  While similar to the $uvby$ system, the $uz,vz,bz,yz$
filters are slightly narrower and the filter response curves are more
symmetrically shaped then the original filters. The $uz,vz,bz,yz$ system
covers three regions in the near-UV and blue portion of the spectrum that
make it a powerful tool for the investigation of stellar populations in
SSP's, such as star clusters, or composite systems, such as galaxies. The
first region is longward of 4600\AA, where in the influence of absorption
lines is small.  This is characteristic of the $bz$ and $yz$ filters
($\lambda_{eff}$ = 4675\AA\ and 5500\AA), which produce a temperature
color index, $bz-yz$.  The second region is a band shortward of 4600\AA,
but above the Balmer discontinuity. This region is strongly influenced by
metal absorption lines (i.e. Fe, CN) particularly for spectral classes F
to M which dominate the contribution of light in old stellar populations.
This region is exploited by the $vz$ filter ($\lambda_{eff} = 4100$\AA).
The third region is a band shortward of the Balmer discontinuity or below
the effective limit of crowding of the Balmer absorption lines.  This
region is explored by the $uz$ filter ($\lambda_{eff} = 3500$\AA). All the
filters are sufficiently narrow (FWHM = 200\AA) to sample regions of the
spectrum unique to the various physical processes of star formation and
metallicity.

Originally, the $uz,vz,bz,yz$ system was convolved with model atmospheres
and Galactic standards to determine its behavior as a function of stellar
type and metallicity.  Later, comparison to stellar population models
(Arimoto \& Yoshii 1986, Guiderdoni \& Rocca-Volmerange 1987, Bruzual \&
Charlot 2000) was used to set the zeropoints for our color system.  All
SED models are presented as a set of spectral energy distributions listing
flux (ergs cm$^{-2}$ s$^{-1}$ Angstrom$^{-1}$) as a function of
wavelength.  Our $uz,vz,by,yz$ colors are determined by convolving the
SED's to our filter sensitivity curves (Rakos \& Schombert 1995).  As an
example, two SED models of $Z$=0.0001 and $Z$=0.02 (where solar [Fe/H]
corresponds to $Z$=0.02) and our filter bandpasses are shown in Figure 1
for a SSP of age 13 Gyrs (the SED's have been normalized at 5500\AA).  The
known trend of bluer colors for the metal-poor populations is obvious.
Note also that the bluer population for the metal-poor model changes the
general shape of the SED and that the increase in the strength of metal
absorption features sharpens the 4000\AA\ break in the metal-rich model.
This is an important point since any interpretation of a color system
requires calibration of the effects of line blanketening as well as the
changes in the effective temperature of the stellar population.  In
addition, the effects of luminosity weighted fluxes from different
metallicity populations will produce significant changes in the integrated
colors (see \S 2.3).  

The Guiderdoni \& Rocca-Volmerange models were used to trace the color
evolution of ellipticals out to redshifts of 0.9 (Rakos \& Schombert 1995)
and the analysis was a good example to the interplay between models and
observations.  For example, the data on ellipticals indicated a red `bump'
at redshifts of 0.4.  This bump is visible in both the continuum $bz-yz$
and metallicity $vz-yz$ indices.  The Guiderdoni \& Rocca-Volmerange
models, which incorporated the details of the late stages of stellar
evolution (HB and AGB populations), predicted this color bump as due to a
8 Gyr old HB contribution that peaked at that particular epoch (on the
assumption of a galaxy formation epoch at a redshift of 5).  However, the
metallicity colors ($vz-yz$) for the bump and more distant clusters were
bluer than Guiderdoni \& Rocca-Volmerange model predictions.  That
difference was due to the lack of a variation in metallicity in the
stellar templates used to produce the SED models.  The models of Arimoto
\& Yoshii corrected this deficiency and predicted the bluer $vz-yz$ colors
in a burst population.  The integrated light from present-day ellipticals
is primarily from red giant branch (RGB) and turnoff main sequence stars
(TO), but in the past (even as recent as 4 Gyrs ago), the contribution
from metal-poor HB and AGB stars increases to a noticeable fraction of the
total light.  Thus, the color evolution of ellipticals demonstrated the
need for models which not only contained the details of the late stages of
stellar evolution plus also incorporated a full treatment of chemical
evolution that follows the extended burst of star formation that created
the first generations.

\begin{figure}
\plotfiddle{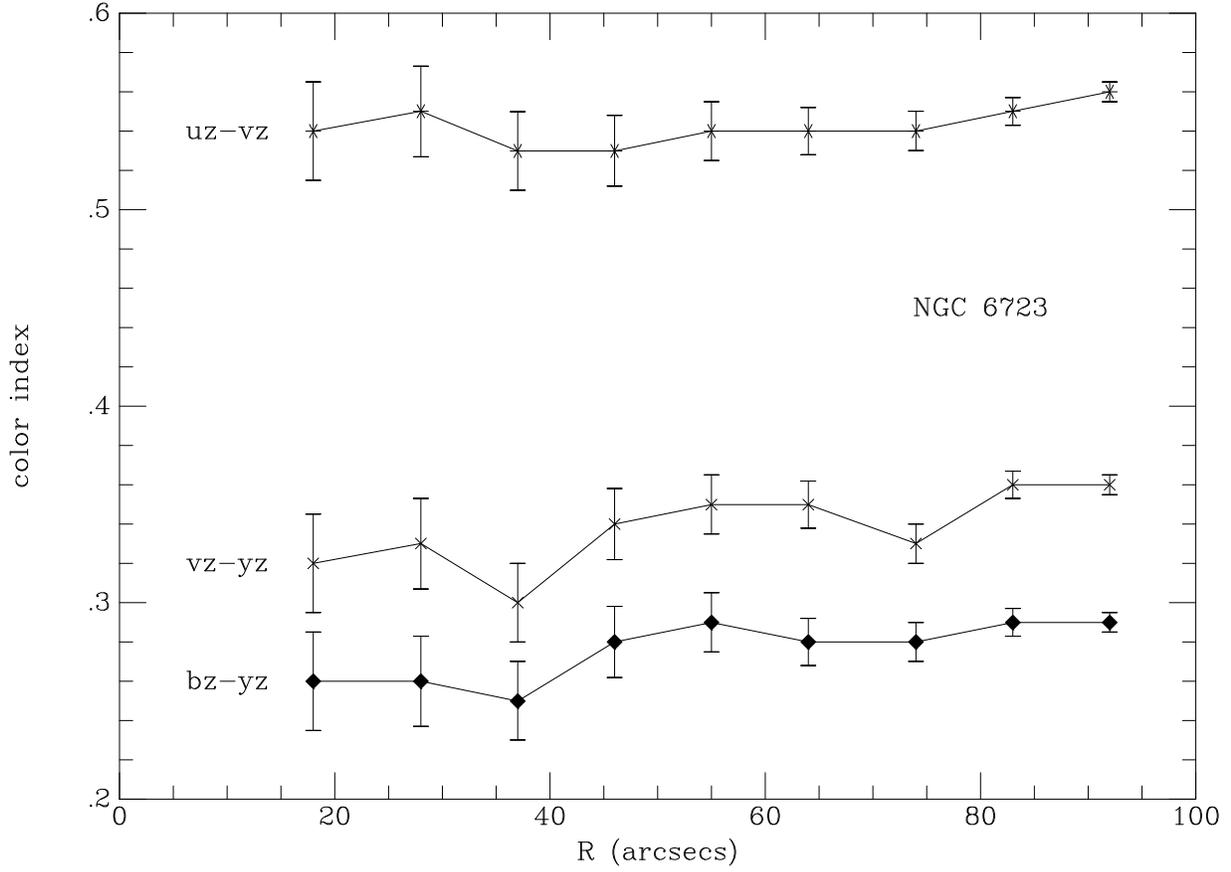}{9.0truecm}{-90}{65}{65}{-270}{360}
\caption{Aperture colors as a function of radius for globular
cluster NGC 6723.  While there is a mild redward color gradient with
radius, colors level off by the half-light radius (85 arcsecs).  Formal
errors range from 0.005 to 0.015 mags 
depending on apparent luminosity.}
\end{figure}

The $uz,vz,bz,yz$ system is designed to match the rest frames of distant
clusters so as to minimize k-corrections.  However, in a parallel study to
the distant ellipticals, the zero-redshift filters were applied to a set
of nearby, bright ellipticals to search for color gradients (Schombert
\etal 1993).  Color gradients reflect either age or metallicity changes
during the formation of a galaxy.  Current studies strongly indicate that
gradients are primarily metallicity induced (Kodama \& Arimoto 1997), but
the Schombert \etal data suggested a problem in the $uz,vz,bz,yz$ system
calibration to Mg$_2$ due to possible light element abundance enhancement
in ellipticals (Worthey 1998).  To investigate this possibility, we have
undertaken a program to relink the $uz,vz,bz,yz$ system to the metallicity
of globular clusters, particularly since a study of low mass galaxies will
enter the realm of less than solar metallicities that few models have
attempted.

\subsection{Globular Clusters}

In order to relink the $uz,vz,bz,yz$ system to the metallicity system defined
by globular clusters, a series of metal-rich and metal-poor systems were
observed between 1997 and 1999 from both the northern and southern
hemisphere.  The northern cluster observations were made at Lowell
Observatory using 1.1m Hall telescope.  The imaging device was the SIT 2K
CCD with 24 micron pixels binned 4x4.  The field size covers 18x18 arc
minutes of sky.  The southern cluster observations were made at CTIO using
the 0.91m Curtis Schmidt telescope plus STIS 2K CCD with 21 micron
pixels.  The plate scale of 2.1 arcsec/pixel covers 76x76 arcmins of sky.
Reduction followed standard CCD procedures (bias, flatfield, etc).
Standard spectrophotometric stars were used for calibrations following the
same procedures for our distant galaxy program (see Fiala, Rakos \&
Stockton 1986, Rakos, Fiala \& Schombert 1988).  Reddenings were
calculated according to Rakos, Maindl \& Schombert (1996), taking into
account the $E(B-V)$ values from the literature and compiled by Harris
(1996).  One of the primary goals of the project is to calibrate an age
and metallicity relation from the SSP's in globular clusters.  To this
end, we have placed a special emphasis on selecting a set of clusters with
varying [Fe/H] and age values.  Since age and metallicity are correlated
in our own Galaxy, it was not possible to completely fill the parameter
space of the two.  However, we have attempted to observe a maximum range
in metallicities in halo and disk clusters of similar age.

Since our primary interest is the integrated color of main stellar
population that comprises the globular clusters, special attention was
given to the influence of foreground stars (due to low Galactic latitude
for most of the sample) and to possible color changes with the aperture
radius.  To investigate the aperture effect, plots were made of the
various colors ($uz-vz$, $vz-yz$ and $bz-yz$) as a function of aperture
radius.  One example is shown in Figure 2 for cluster NGC 6723 ($b =
-17\deg$).  Very little change in color is found outside the inner 40
arcsecs of the cluster (the half light radius for cluster is 85 arcsecs).
This confirms previous work that there is little change in the stellar
population of a globular cluster from core to edge in terms of mean color
of the stars.  Experimentation demonstrated that the most stable aperture
is the half-light radius of the cluster.

The influence of the foreground stars is estimated by measuring several
regions on each CCD frame outside of the cluster.  The total difference
between cluster and brightness in the neighborhood was, on average, at
least 5 magnitudes fainter (for the same aperture size) then the cluster
luminosity.  This would account for, at most, 0.010 mag of total light.
Experimentation with sky subtraction produced a variance of only 0.005 mag
in cluster color at the half-light radius.  The accuracy of our final
measurements ranged from 0.005 to 0.015 magnitudes depending on the
brightness of the cluster.  The resulting $uz,vz,bz,yz$ photometry (all
colors refer to half-light values) is listed in Table 1 where column 1 is
the NGC number of the cluster, column 2 is the $uz-vz$ color, column 3 is
the $bz-yz$ color, column 4 is the $vz-yz$ color, column 5 is the $mz$
index, column 6 is the cluster age in Gyrs (Salaris \& Weiss 1998), column
7 is the mean cluster [Fe/H] (Harris 1996, Carretta \& Gratton 1997),
column 8 is the extinction $E(B-V)$ (Harris 1996), column 9 is the HB
index (Lee, Demarque \& Zinn 1994) and column 10 is the $\Delta(bz-yz)$
index (see below).

\begin{figure}
\plotfiddle{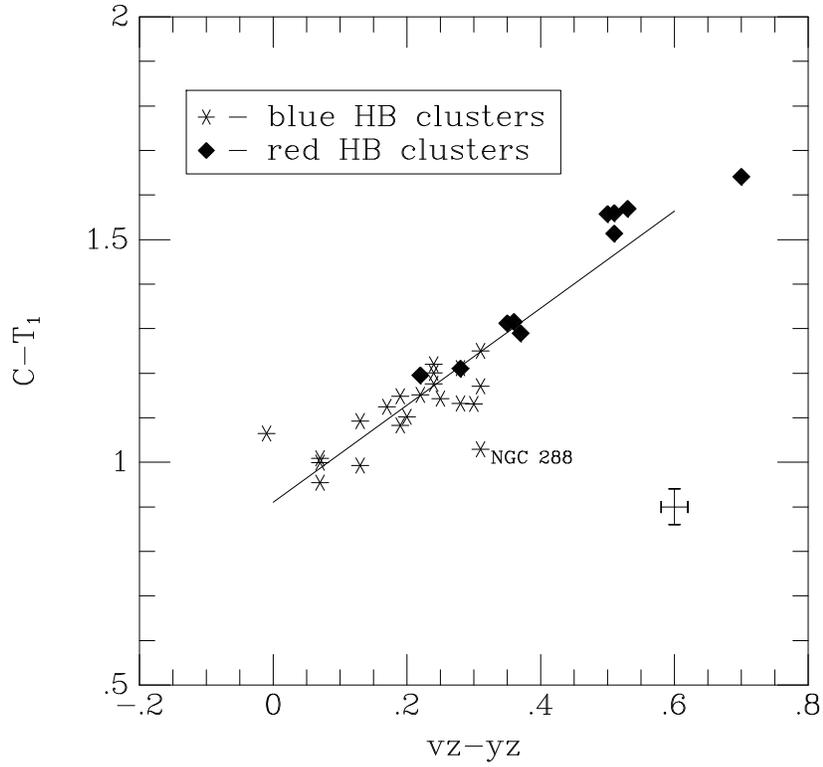}{8.5truecm}{0}{75}{75}{-240}{-180}
\caption{The Washington metallicity index, $C-T_1$ versus our metallicity index,
$vz-yz$ for 32 globular clusters.  The correlation is good, except for the
anomalous cluster NGC 288.  The solid line is a least-squares fit to the
data (excluding NGC 288).  There is no statistical difference between blue
and red HB clusters (the second parameter effect).}
\end{figure}

A similar calibration to globular cluster metallicities was made with the
Washington $CMT_1$ system by Geisler \& Forte (1990).  The Washington
broadband filters covered 3910\AA\ ($\Delta\lambda=1100\AA$), 5085\AA\
($\Delta\lambda=1050\AA$) and 6330\AA\ ($\Delta\lambda=800\AA$).  The
$C-T_1$ index is well calibrated to cluster [Fe/H] (Geisler \& Forte 1990,
Cellone \& Forte 1996)).  From their study of 48 clusters, 30 overlap with
our $uz,vz,bz,yz$ photometry.  Using our $E(B-V)$ values, their
metallicity index $C-T_1$ and our $vz-yz$ index is plotted in Figure 3.
The correlation between our systems is good even though the filter widths
are dramatically different.  This probably reflects the well defined
stellar populations in globular clusters, in terms of a narrow turnoff and
RGB sequences, as they convolve into color indices.
Since the stars in a cluster have a narrow range in both age and
metallicity, the resulting CMD is sharp and the contrast between clusters
is clear to both narrow and broad band photometry.  Notice, also, that
there is no obvious difference between red and blue HB clusters (see
below).  One deviant cluster is NGC 288, a metal-rich system with a
history of difficulty in defining its age from isochrone fits (also see
below).

For calibration to a metallicity system, one would ideally like the total
amount of heavy metals in a system, $Z$.  Traditionally, the field of
chemical evolution has used the ratio of iron and hydrogen normalized to
the solar value, [Fe/H], as a direct tracer of $Z$.  Globular cluster
[Fe/H] measurements, based on isochrone fits, vary from author to author.
While internal errors are quoted from 0.02 to 0.15 dex in the literature,
a comparison between three metallicity systems (Zinn 1985, Rutledge \etal
1997 and Carreta \& Gratton 1997) demonstrates a mean external error of
0.2 dex that we will adopt for our analysis.  The [Fe/H] values used
herein are listed in Table 1.  We have also assumed that the spread of
[Fe/H] within a cluster is less than our observational error, a fairly
safe assumption given the narrowness of the giant branches in Milky Way
clusters.

\begin{figure}
\plotfiddle{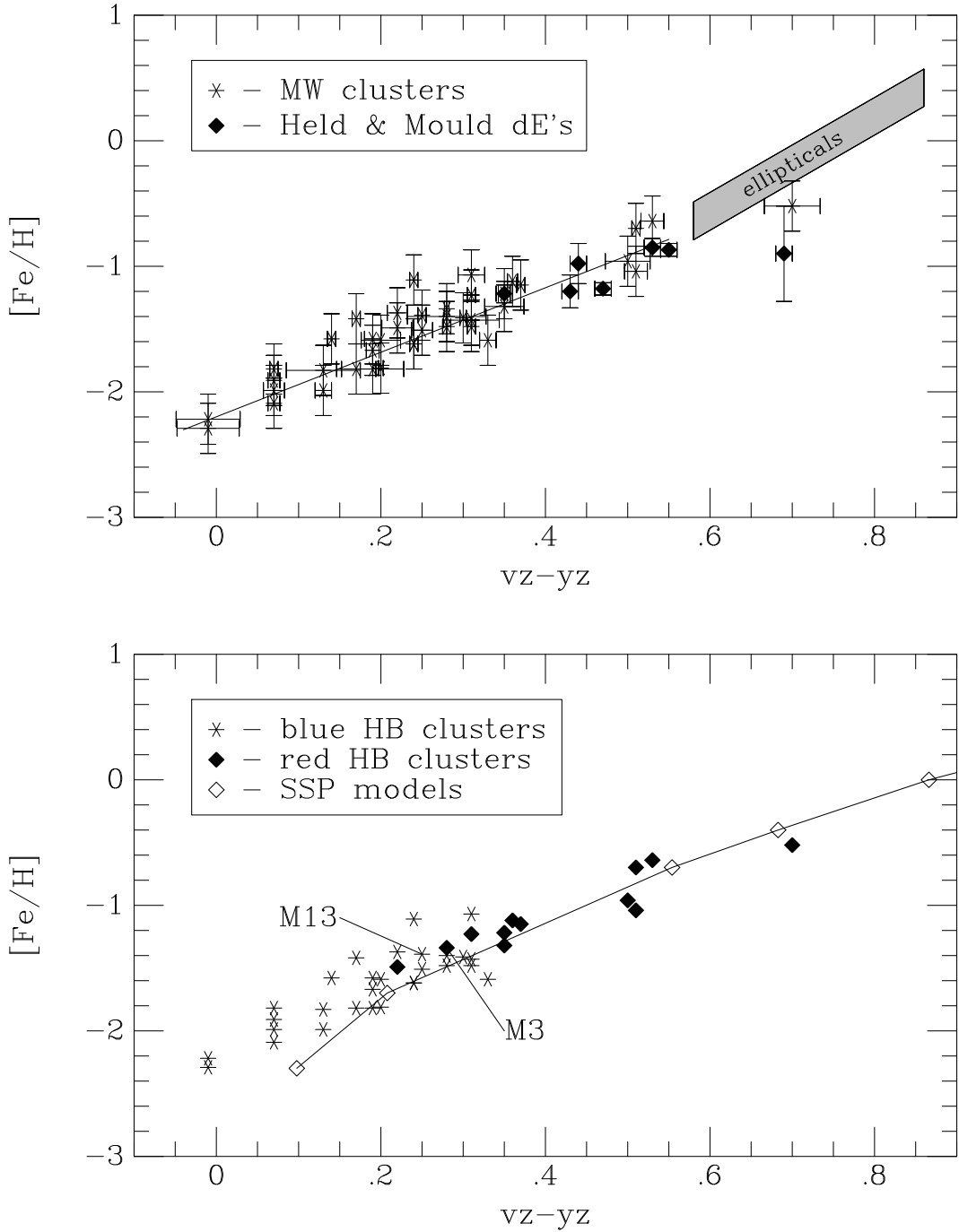}{17.0truecm}{0}{80}{80}{-230}{-50}
\caption{The metallicity color, $vz-yz$, versus [Fe/H] for the 41
globular clusters listed in Table 1 (crosses) and the Fornax dwarf
ellipticals (filled diamonds) with spectroscopic measurements of [Fe/H]
(Held \& Mould 1994).  The shaded region is the color-magnitude relation
from Schombert \etal (1993) calibrated to the Kuntschner (2000)
mass-metallicity relation.  The bottom panel displays the globular cluster data
divided into those with blue and red HB branches (separated by Lee index,
Lee, Demarque \& Zinn 1994).  M3 and M13, two extreme HB morphology
clusters, are indicated.  There is no evidence that the second parameter
effect distorts the metallicity calibration.  The 13 Gyr SSP model of
Bruzual \& Charlot (2000) is also shown.}
\end{figure}

Metallicity in the $uz,vz,bz,yz$ system is given by the $vz-yz$ index
which compares the light from a portion of the spectrum with several
blends of metallicity indicators (Fe, CN) to a continuum region.  The
globular cluster relationship between [Fe/H] and our metallicity index,
$vz-yz$, is shown in the upper panel of Figure 4.  The correlation is
excellent with a formal linear fit given by:

$$ [Fe/H]_{gc} = 2.57\pm0.13(vz-yz) - 2.20\pm0.04 $$

\noindent for the range of [Fe/H] of $-2.5$ to $-0.5$.  There is a small
gap around the [Fe/H] = $-1.0$ region, of critical importance for studying
dwarf ellipticals.  However, that region can be filled with [Fe/H]
values determined spectroscopically for Fornax dE's (see \S2.3).

\begin{figure}
\plotfiddle{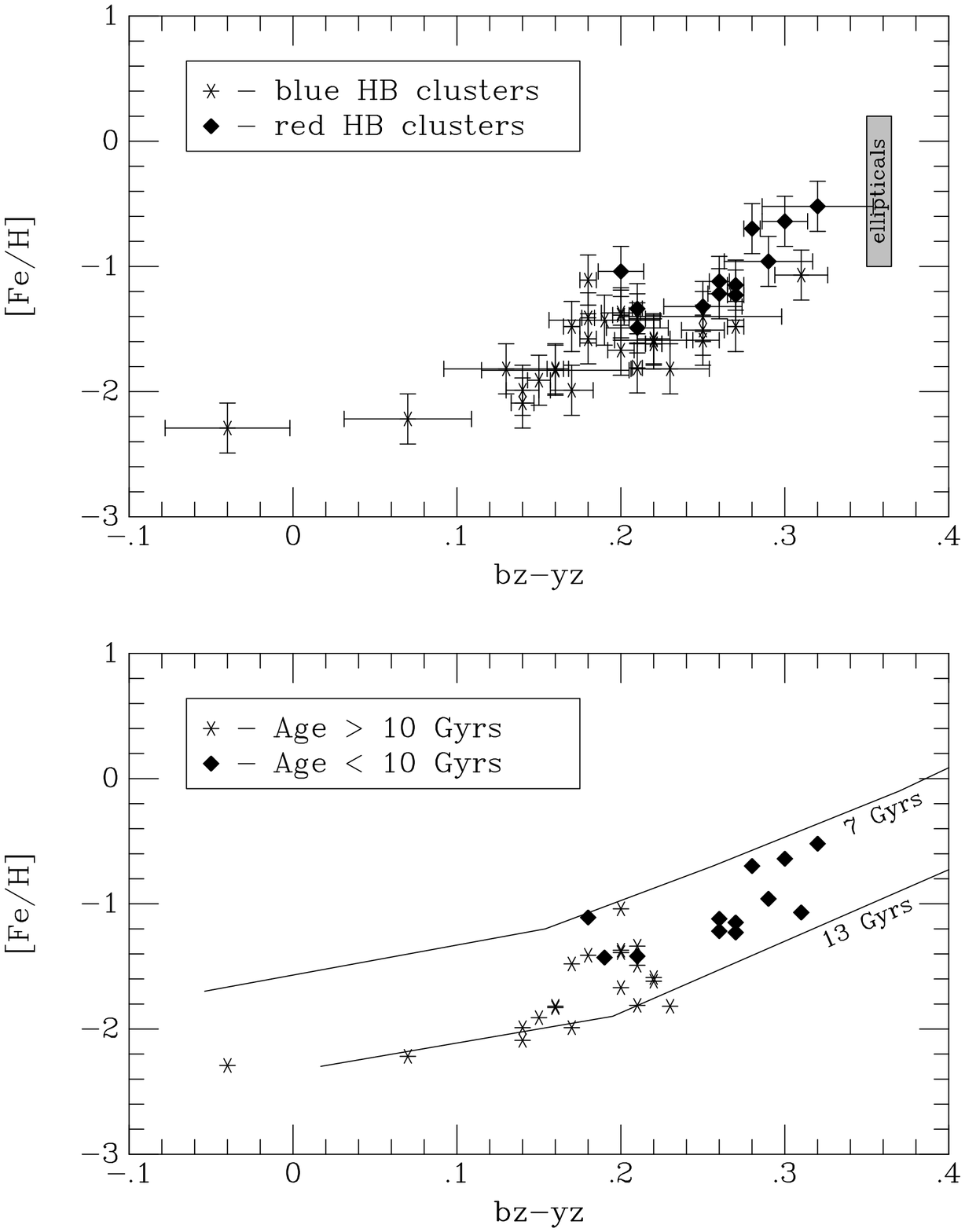}{17.0truecm}{0}{80}{80}{-230}{-50}
\caption{The continuum color, $bz-yz$, versus metallicity [Fe/H] for
the Milky Way globular clusters.  The top panel displays the cluster data
divided by HB morphology as per Figure 4.  The bottom panel displays the
data for those clusters with isochrone ages (Salaris \& Weiss 1998) along
with two model tracks at 7 and 13 Gyrs from Bruzual \& Charlot (2000).
The shaded box in the top panel represents the run of $bz-yz$ color and
metallicity for bright ellipticals.}
\end{figure}

The correlation in Figure 4 is not due solely to line blanketening effects
as would be the case for individual stars.  There is a well-known shift in
the effective temperature of the stars that make up the TO and RGB of the
clusters as global metallicity of the stellar population vary.  For this
reason, the continuum index, $bz-yz$, also varies with metallicity, but
with much greater scatter (shown in Figure 5) indicating the importance of
the line blanketening component even in composite stellar systems.
We note that the $vz-yz$ index varys both from temperature and line
blanketening effects which combine in the same direction (more flux in the
$vz$ relative to $yz$ as the mean metallicity decreases) to tighten the
correlation.

Of serious concern is the effect of age and horizontal branch morphology
on the metallicity calibration.  The location of HB stars in the
color-magnitude diagram (CMD) can vary from a position on the extremely
blue edge of the instability strip (e.g. M13) to a red clump against the
base of the giant branch (e.g. M3).  Since M3 and M13 have similar
metallicities, then some other parameter influences HB morphology (the
so-called second parameter problem).  To first order, the addition of
another blue component would drive the $vz-yz$ metallicity estimate to
lower [Fe/H] values.  And, since [Fe/H] and HB morphology are correlated
in the same direction (i.e. bluer population for lower metallicity), then
it is possible that the metallicity calibration is distorted by HB
morphology and the calibration would disappear in populations with a
mixture of blue and red HB stars.

To determine the effect of HB morphology on the metallicity calibration,
we have plotted blue and red HB clusters in the bottom panel of Figure 4.
To make this division, we have assigned clusters with HB ratios, given by
$(B-R)/(B+V+R)$ (Lee, Demarque \& Zinn 1994), of less than 0.1 the
designation of red and those with HB ratios greater than 0.1 the
designation of blue.  The first parameter effect, that red HB clusters
have higher metallicities, is obvious.  In the overlap region, there is no
trend to separate the sequences by $vz-yz$ color.  Even M3 and M13, the
extreme examples of HB morphology, are located within their photometric
errors of each other on the calibration line.  A similar lack of
separation by HB morphology in $bz-yz$ as shown in Figure 5.

The lack of a second parameter effect to the $vz-yz$ index is surprising
since the $uz,vz,bz,yz$ system is centered in the blue portion of the
spectrum and should be sensitive to effects of bright, blue stars.  There
are two possible competing effects to the HB morphology that minimizes
their influence on the metallicity index, $vz-yz$.  One is that blue HB
stars are extremely hot and emit most of their energy in the UV.  Their
spectral energy distribution across the $vz$ and $yz$ spectral regions are
much less than expected for their bolometric luminosities.  Thus, clusters
with a developing blue HB component simply contribute less light to the
$vz-yz$ as red HB clusters.  This dropout of blue HB clusters is also seen
in the Washington $CMT_1$ photometry system (Geisler \& Forte 1990).

Secondly, and perhaps more important, is that the $uz,vz,bz,yz$ filters
are dominated by main sequence and RGB stars.  The contribution from HB
stars is less than 5\% of the integrated light for an old stellar
population (Guiderdoni \& Rocca-Volmerange 1987).  Based on comparison to
SSP models, changes in the turnoff point and RGB populations, even small
ones, have a larger effect on $uz,vz,bz,yz$ colors than the HB morphology.
The result is that metallicity calibration, to first order, appears free
from distortions due to blue HB stars.

Changes in the age of a SSP are reflected in the location of both the TO
and RGB populations in the CMD.  However,  the RGB shifts only slightly
compared to the color of the TO population due to the relative
contributions of the TO and RGB to near-blue portion of the spectrum (see
Stetson, Vandenberg \& Bolte 1996).  Calibration with age is more
difficult with photometric systems due to the well known age-metallicity
degeneracy since an age change produces an identical change in colors or
spectral indices if the $\Delta$log Age/$\Delta$log$Z$ = $-3/2$ (Worthey
1994).  With respect to the $uz,vz,bz,yz$ system, age effects at constant
metallicity reflect changes in the turnoff point of the composite CMD
relative to the RGB contribution.  In particular, for constant
metallicity, the RGB changes only slightly with age (blueward) compared to
relatively large changes in the color of the turnoff stars.  Our method
for resolving the age-metallicity degeneracy is to determine the expected
$bz-yz$ color from the $vz-yz$ color (since age has little effect on the
$vz-yz$ metallicity calibration) and correlate the residuals from expected
and observed $bz-yz$ with age.

This age effect can be partially seen in the bottom panel of Figure 5, the
plot of $bz-yz$ versus [Fe/H].  The scatter with metallicity is much
higher compared to $vz-yz$.  The bottom panel plots the expected $bz-yz$
color versus metallicity for a 7 and 13 Gyrs population from Bruzual \&
Charlot models.  Even though the isochrone ages have at least 0.5 Gyrs of
internal error, it is clear that much of the scatter in Figure 5 is due to
an age difference (younger clusters being too blue for their [Fe/H]
values).  While this limits the use of the $bz-yz$ index for metallicity
calibration, it does open up the possibility of using the residuals in
$bz-yz$ to estimate age.

The first step in an age calibration is the correlation of $bz-yz$ with
$vz-yz$.  This correlation is good and connects with the previous
correlation of ellipticals (Rakos, Maindl \& Schombert 1996).  There is
more scatter in $bz-yz$ with respect to the metallicity of the cluster,
but this is expected since the $bz$ and $yz$ filters are in a featureless
portion of the spectrum and correlate with metallicity only as it changes
the mean effective temperature of the underlying stellar population.  Our
procedure, once the correlation between $vz-yz$ and $bz-yz$ is known, is
to use the knowledge provided by the metallicity index, $vz-yz$, which is
relatively free from age and second parameter effects, to fix the
metallicity color of an object.  Then, we use the correlation between
$vz-yz$ and $bz-yz$ to predict the expected $bz-yz$ index for the
metallicity in question.  Any residuals between the expected $bz-yz$ color
and the actual observed $bz-yz$ color should measure changes due solely to
age of the underlying population (or recent star formation, which we
ignore for the present).

\begin{figure}
\plotfiddle{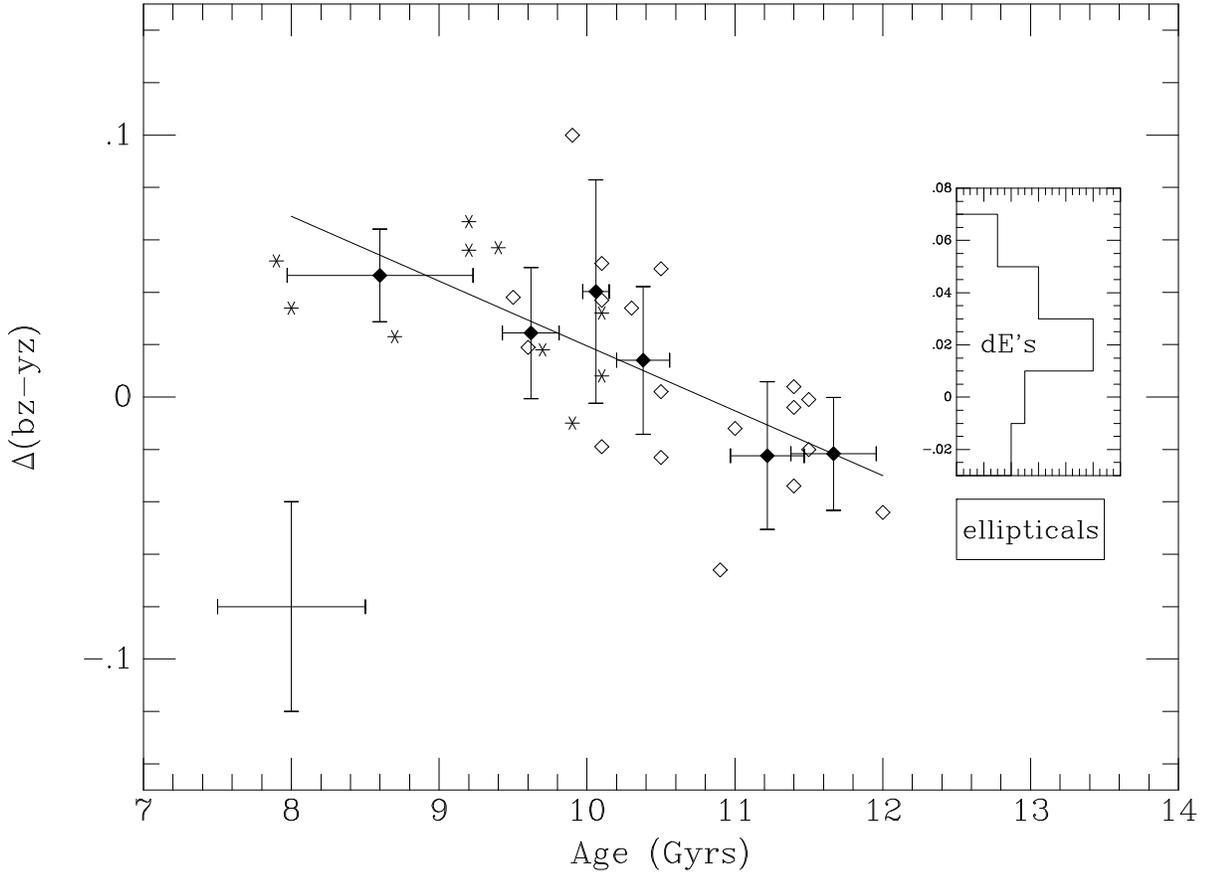}{9.0truecm}{-90}{65}{65}{-270}{360}
\caption{The residual color age indicator, $\Delta(bz-yz)$, versus age in
Gyrs.  The raw globular cluster data is divided by HB morphology from blue
(crosses) to red (open diamonds).  Ages are from Salaris \& Weiss (1998).
A typical error for the globular colors and ages is shown in the bottom
left corner.  Average values (filled diamonds) are displayed along with a
linear fit.  The range for bright ellipticals ($M_B > -18$) is indicated.  
An inset histogram of $\Delta(bz-yz)$ values for
the Fornax dwarfs is also shown.  The $\Delta(bz-yz)$ values for bright
ellipticals indicate a mean age of 13 Gyrs; whereas, the dwarfs have a
mean age of 10 Gyrs.}
\end{figure}

The globular cluster residuals in $bz-yz$, designated $\Delta(bz-yz)$, are
plotted in Figure 6 with respect to cluster age.  For this study, we have
used the new age determinations presented by Salaris \& Weiss (1998).
These new age determinations incorporate the latest changes in stellar
physics, such as the equation of state and opacities, and also use the
HIPPARCOS-based distances to the clusters.  Their new calibration finds
the age of the oldest globular clusters to be approximately 12 Gyrs, in
line with determinations of the age of the Universe from cosmological
constants.  Comparison of age values from other work (Stetson \etal 1999,
Carretta \etal 2000) requires the addition of 4.5 Gyrs to Salaris \& Weiss
values.

For clarity, the age data for clusters from Table 1 have been divided into
six bins of five clusters each.  The error bars in Figure 6 display the
range of $\Delta(bz-yz)$ and age.  The correlation evident in Figure 6 is
weak, mostly due to combined errors in the age determination and
photometric corrections, but sufficient to estimate SSP ages at the 1 Gyr
level and results in the following 
calibration:

$$ T_o = -33.97 \times \Delta(bz-yz) + 10.72 $$ 

\noindent where $T_o$ is the age of the cluster in Gyrs.  Note that the
anomalous cluster NGC 288 is dropped from the age analysis because it
displays discrepant age indicators.  While Salaris \& Weiss determine NGC
288's age to be 8.8 Gyrs, other determinations (based on isochrone fits)
are 2 to 3 Gyrs older (e.g. Alcaino, Liller \& Alvarado 1997).  In
addition, NGC 288 is highly discrepant on the Washington calibration
system, (Geisler \& Forte 1990), again suggesting a calibration problem
for the cluster.  The $\Delta(bz-yz)$ index indicates an age of 10 to 11
Gyrs for NGC 288, more in line with other CMD ages; however, due to its
poor isochrone fits, it was dropped from the above analysis.  

\subsection{SSP Models}

Globular clusters provide the simplest test to the SSP models since their
stars are restricted in age and metal content.  Galaxies, on the other
hand, form their stars over some duration of time with the metallicity
increasing for each new generation.  For galaxies, the simplest model is one 
the total stellar population is composed of a series of SSP's assuming
some metallicity distribution based on a chemical evolution scenario
(Greggio 1997) and a star formation history (usually a single burst of a
fixed duration for ellipticals).  In order to investigate the effects of a
composite stellar population with a range of metallicities (and ages) on
the behavior of our color indices, we have adopted the SSP models of
Bruzual \& Charlot (2000, hereafter BC2000) with a Miller-Scalo IMF for
the range of 0.1 $M_{\sun}$ to 100 $M_{\sun}$.  The models follow a
single-burst stellar population over a range of ages (1 to 17 Gyrs) with a
range of metallicities (1/200 to 2.5 times solar).  A full description of
the details of the modeling can be found in Liu, Charlot \& Graham
(2000).

The BC2000 models are presented as a set of spectral energy distributions
listing flux (ergs cm$^{-2}$ s$^{-1}$ Angstrom$^{-1}$) from 91\AA\ to
16,000\AA.  Our $uz,vz,by,yz$ colors are determined by convolving the
SED's to our filter sensitivity curves (Rakos \& Schombert 1995).  At any
particular epoch, the integrated SED of a galaxy will be composed of a
series of populations of different metallicities and ages.  What is
observed will be the fluxed-weighted sum of the different populations.
Because of this effect, the [Fe/H] value determined from the ($vz-yz$)
color will be in error if the distribution of real metallicities is not
gaussian.  In order to determine the effect of a luminosity weighted
metallicity, a particular model of chemical evolution is required.  To
this end, we have used the simple (versus infall) model of Kodama \&
Arimoto (1997).  This model assumes a closed box scenario (e.g. Gibson \&
Matteucci 1997) where the gas is well-mixed and uniform.  The gas is
enriched by ejection of metals from dying stars and succeeding populations
are born from the enriched gas.  This process continues until some event,
such as galactic winds, halts the star formation process.  While there may
be complicating factors (such as initial enrichment or infall of gas-poor
clouds), this model provides the most direct test of our assumptions and
we will argue that changes in the basic assumptions make little difference
to the resulting colors.

\begin{figure}
\plotfiddle{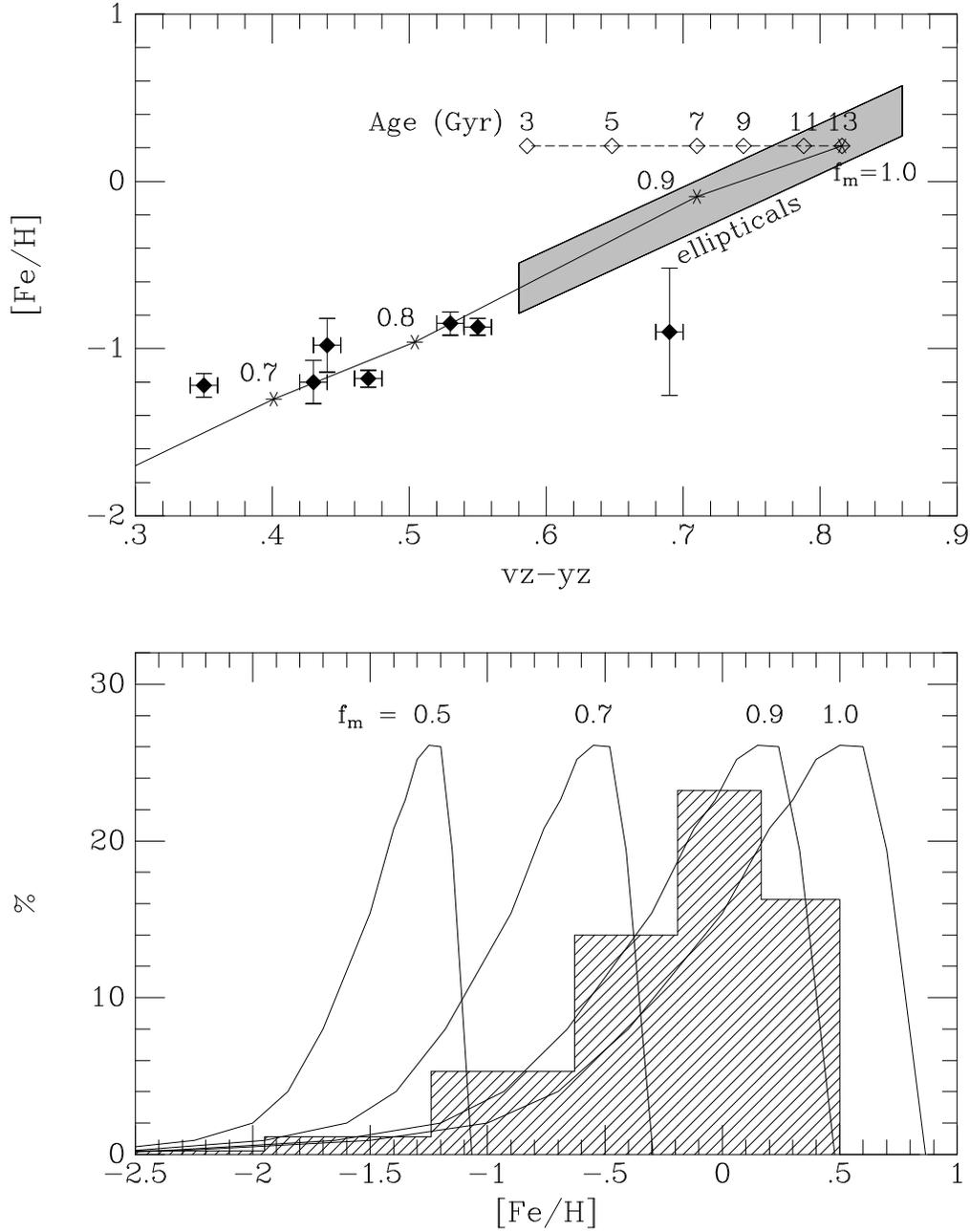}{16.0truecm}{0}{75}{75}{-230}{-40}
\caption{Multi-metallicity populations in $vz-yz$, [Fe/H] space.  The bottom panel
displays the Kodama \& Arimoto (1997) metallicity distribution for a
standard elliptical model ($f_m=1.0$).  Lower metallicity models are
produced by reducing the standard elliptical model by a constant ($f_m$).
The quantization to the BC2000 models ([Fe/H] = $-$2.3, $-$1.7, $-$0.7,
$-$0.4, 0.2 and 0.5) is shown as the shaded histogram for the $f_m=0.9$
model.  The upper panel shows the resulting $vz-yz$ colors and luminosity
weighted mean metallicity for the various $f_m$ values.  The shaded region
is the color-metallicity relation from Schombert \etal (1993) using the
Kuntschner (2000) calibration.  The Held \& Mould (1994)
spectroscopic data on dwarf ellipticals (filled diamonds) is also shown.
In addition, the dashed line shows the $f_m=1.0$ model for various ages.  The
multi-metallicity models are an excellent fit to the bright and dwarf
elliptical data, in contrast to the clear discrepant nature to the bright
ellipticals from the globular cluster relation in Figure 4.}
\end{figure}

The bottom panel of Figure 7 displays the metallicity distribution
required to fit the mass-metallicity sequence of ellipticals (Kodama \&
Arimoto 1997).  The simple model results in a distribution of stars per
metallicity given in Table 2.  It has a peak at [Fe/H]=0.5, but the
numerical mean is 0.21 and, most importantly, the luminosity weighted [Fe/H]
value is $-$0.10.  To simulate a sequence of decreasing metallicity in
galaxies, we have compressed the Kodama \& Arimoto distribution by
fractions, $f_m$, from the peak value of $+$0.5 dex (mean [Fe/H]=0.21).
Computationally, the lowest bin is fixed at [Fe/H]=$-$2.3 and the
metallicity distribution is multiplied by $f_m$.  This has the effect of
lowering the mean metallicity while maintaining the shape of the
distribution.  We prefer this method, versus simply shifting the
metallicity distribution to lower values, since it maintains the
sparseness of metal-poor stars per unit volume (the G-dwarf problem) as is
observed in real galaxies (Worthey, Dorman \& Jones 1996).  Three such
compressed distributions are shown in Figure 7 and listed in Table 2.  The
resulting average metallicity of these distributions is given by the
formula $<[Fe/H]> = 5.51 f_m - 5.23$.  Since the Bruzual \& Charlot models
are sorted in [Fe/H] bins of $-$2.3, $-$1.7, $-$0.7, $-$0.4, 0.0 and
$+$0.4, the Kodama \& Arimoto distribution is quantized by those values
(shown as the shaded histogram for $f_m=0.9$ in Figure 7).

For each generated metallicity distribution, the appropriate population
from the BC2000 models are assigned, weighted and summed to produced the
total flux for the integrated population.  The resulting $vz-yz$ colors
for a 13 Gyr population are shown in the top panel along with the
color-magnitude relation for bright ellipticals (in this paper we refer to
bright ellipticals as those galaxies with total luminosities greater than
$L^*$) and data from Held \& Mould (1994) for seven Fornax dE's where
there is $vz-yz$ data to match the spectroscopically determined [Fe/H]
values.  The color-magnitude relation will be discussed in greater detail
in \S 2.5.  In brief, the observed color-magnitude relation in $vz-yz$ is
converted to the [Fe/H] scale using the relation of magnitude-[Fe/H] from
Kuntschner (2000).  Figure 7 confirms that the simple metallicity
model outlined above precisely reproduces the $vz-yz$ versus metallicity
relationship for globular clusters, dwarf galaxies and bright
ellipticals.

\begin{figure}
\plotfiddle{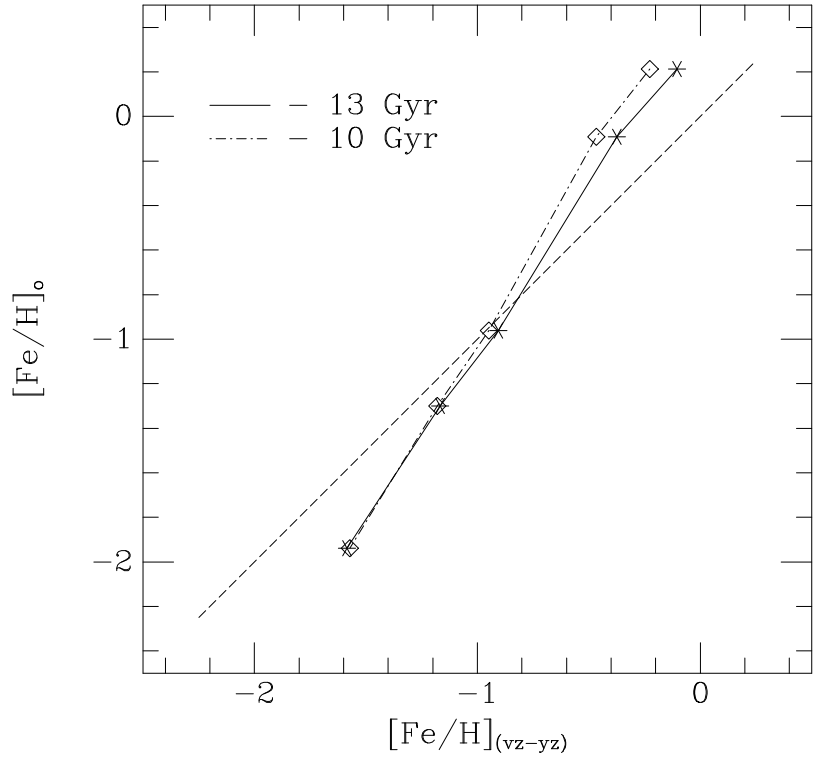}{8.0truecm}{0}{75}{75}{-230}{-180}
\caption{The observed [Fe/H] value versus the true [Fe/H] from
the multi-metallicity models.  The correction curve is shown for two ages,
10 and 13 Gyrs.  In general, the [Fe/H] value determined from $vz-yz$
based on the globular cluster calibration will underestimate the true
[Fe/H] due to contributions from stellar populations with a range of
metallicities above [Fe/H] = $-1$ and overestimate below $-$1.  The effect
of age on the corrections is minor.}
\end{figure}

The result of the model comparisons is that the raw $vz-yz$ color of a
galaxy with a composite stellar population will underestimate the real
metallicity if the globular cluster calibration is applied without
correction for the metallicity distribution of the underlying stellar
population.  For example, the $vz-yz$ color of the $f_m=1.0$ model would
have produced a luminosity weighted [Fe/H] from the globular cluster
calibration of $-$0.10 dex, whereas the real mean [Fe/H] value was $+$0.2
dex.  In order to determine the mean metallicity of a galaxy, we must
correct the luminosity-weighted color as guided by the model results.
This multi-metallicity correction is shown in Figure 8 for both 13 and 10
Gyr models.  There is only a minor difference as a function of age, so we
adopt a linear fit to the 13 Gyr correction such that

$$ [Fe/H]_{o} = 3.78\pm0.14(vz-yz) - 2.83\pm0.04 $$

\noindent where [Fe/H]$_o$ is the mean metallicity of the entire population.
We have also examined other metallicity distributions, such as the infall
model of Kodama \& Arimoto (1997), and they produce negligible differences
in our results.

\subsection{Fornax Dwarf Ellipticals}

The observations of the Fornax dwarf ellipticals were made at CTIO using
the 0.91m Curtis Schmidt telescope plus STIS 2K CCD with 21 micron pixels.
The plate scale of 2.1 arcsec/pixel covers 76x76 arcmins of sky.  The dE's
were selected from Ferguson catalog with high probability of membership
based on morphological criteria (Ferguson 1989).  While membership based
on morphology may be questionable for most Hubble types, dwarf ellipticals
have a particular appearance on photographic material (low surface
brightness, exponential luminosity profile) that easily distinguishes them
from background ellipticals.  Membership is found to be good to 98\% in
both Virgo and Fornax using this method (Ferguson 1994).  Two examples
from our sample are shown in Figure 9, FCC 116 and FCC 207.  They cover
the range of surface brightness typical for our sample, although there
exist many dwarf ellipticals with much lower surface brightnesses and the
sample is not complete with respect to luminosity density.

\begin{figure}
\plotfiddle{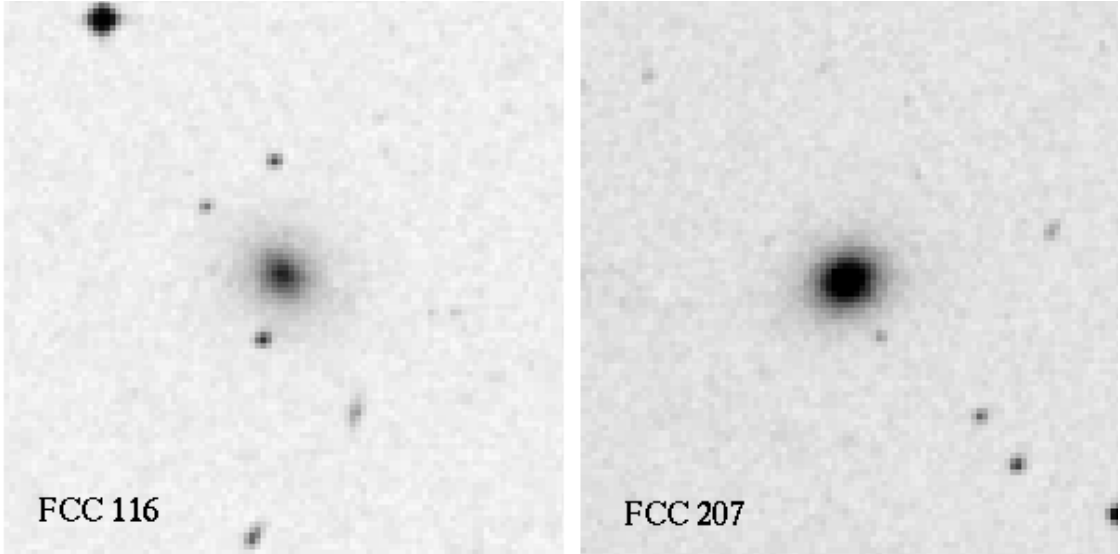}{6.5truecm}{0}{70}{70}{-210}{-180}
\caption{Two examples of the Fornax dwarf ellipticals in the sample. On
the left is FCC 116, a low surface brightness dwarf, and on the right FCC
207, a high surface brightness dwarf.  Each image is 3 arcmins in width.}
\end{figure}

Reduction and calibration followed the same procedures used for the
globular clusters.  Individual exposures for the dE's were long (typically
300 to 600 secs) and coadded to remove cosmic ray artifacts.  Due to the
plate scale of the Curtis Schmidt, none of the dE's were more than 5 to 10
pixels in radius which prevented a full surface photometry analysis of
their colors.  Values quoted are based on aperture magnitudes of 16 kpc
metric size which is sufficient to contain the entire visible area of the
galaxy.  The reddenings of dwarfs were calculated according to Rakos,
Maindl \& Schombert (1996) and assuming $E(B-V)=0.03$ for the Fornax
cluster (Hanes \& Harris 1986). Results for the observed 27 dE's are
listed in Table 3.  The distance to Fornax assumed to be 18.6 Mpc (Madore
\etal 1999).

All the dE's selected for this study are of the nucleated type (dE,N, see
Sandage, Binggeli \& Tammann 1985).  This was due primarily to the need to
compare our photometry with [Fe/H] values determined spectroscopically in
dE's.  Due to their low surface brightness nature, only nucleated objects
have been subjected to spectroscopy.  While this may bias our results if
the existence of a nucleus is an indication of recent star formation,
broadband color studies indicates no difference between the color of the
nuclei and the envelopes of dwarfs (Caldwell \& Bothun 1987).  The central
peak appears to be of dynamical origin rather than a star formation
event.  In addition, dE's exhibit very little in the way of color
gradients (Bothun \& Mould 1988) that might distort our color indices with
respect to mean metallicity.

A check of our globular cluster metallicity calibration directly to dwarf
ellipticals is possible through the spectroscopy data from Held \& Mould
(1994).  There are seven objects with both spectroscopy and $uz,vz,bz,yz$
photometry (FCC 85, 150, 188, 207, 245, 266 and 296). All seven are
plotted in Figure 4 using the median [Fe/H] values from Table 4 of Held \&
Mould.  Notice that the Held \& Mould dE's conveniently cover the
intermediate metallicity range between metal-poor halo globular clusters
and metal-rich disk clusters.  The dE $vz-yz$ colors are also in the
region where the multi-metallicity corrections from the equations in \S2.3
are small.  All the Held \& Mould dE's follow the same relationship as the
globular clusters, within their errors, even without the multi-metallicity
corrections discussed in \S 2.3.  The reddest dE (FCC 85) has the largest
[Fe/H] uncertainty, but still falls on the cluster sequence within its
error.  From the fact that the global $vz-yz$ and spectroscopic [Fe/H]
values are a good fit to the globular cluster relation in Figure 4, we
conclude that total color measurements are in agreement with nuclear
spectroscopy.

\begin{figure}
\plotfiddle{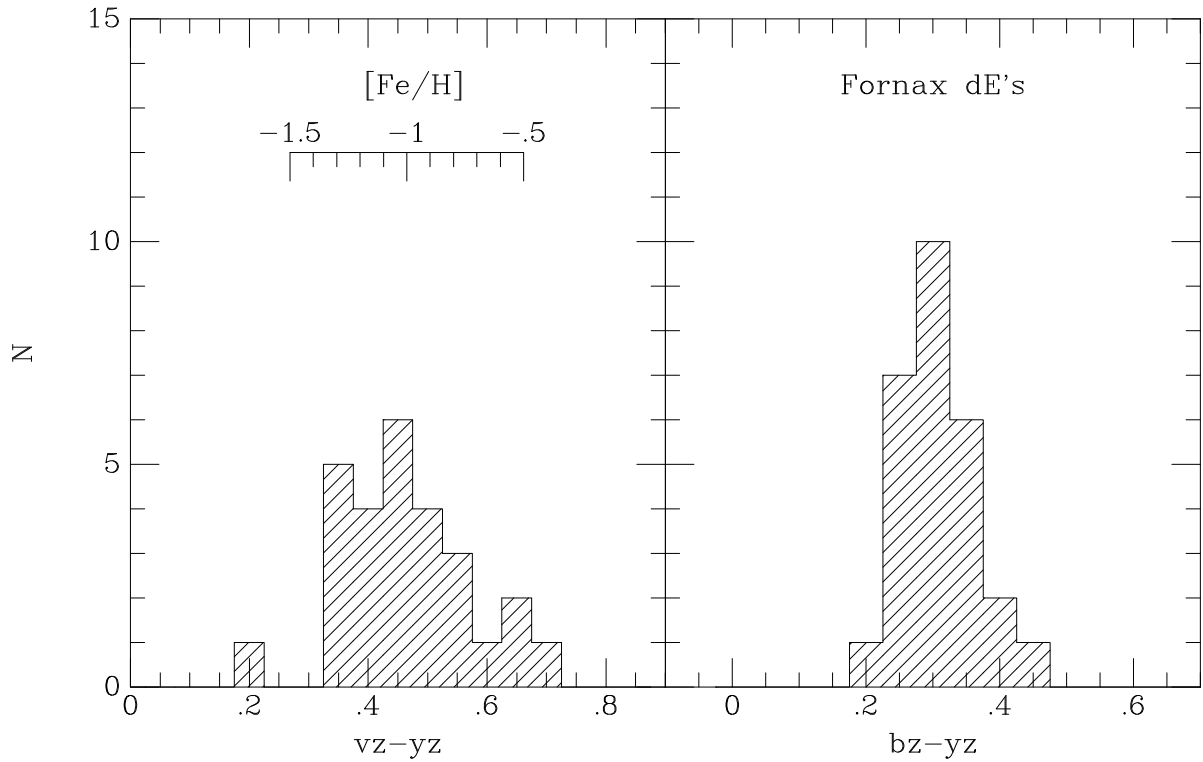}{10.0truecm}{-90}{70}{70}{-270}{350}
\caption{Histograms of $vz-yz$ and $bz-yz$ for the Fornax dwarf
ellipticals.  The calibration from $vz-yz$ to [Fe/H] from equations in \S
2.4 is also shown.  Dwarf ellipticals have a mean [Fe/H] of $-$1.0 ranging
from $-$1.6 to $-$0.4.  The mean $bz-yz$ color is $0.30\pm0.05$.}
\end{figure}

The distribution of $vz-yz$ and $bz-yz$ colors for the Fornax dE's can be
seen in the histograms of Figure 10.  The mean $vz-yz$ color is
$0.46\pm0.11$ and the mean $bz-yz$ color is $0.30\pm 0.05$.  The $vz-yz$
colors transform into a corrected mean [Fe/H] value of $-1.00\pm0.28$;
however, the actual range of metallicities is from moderately metal-poor
([Fe/H]=$-$1.6) to values similar to metal-rich disk cluster
([Fe/H]=$-$0.4) .  This confirms what was found in both optical, IR and
spectroscopy studies of dwarf ellipticals, that their metallicities are
less than that of bright ellipticals and the disk of our own Galaxy, but
greater than that of halo clusters (Bothun \& Mould 1988).

The age of dwarf ellipticals can be estimated using the $\Delta(bz-yz)$
index discussed in \S 2.1.  Note that the age of composite SSP's, such as
an elliptical galaxy, is the luminosity-weighted mean age of entire
stellar population.  If the star formation rate was proportional to the
gas supply, then we would expect the SFR to be smooth function and the
luminosity-weighted mean age to represent the time from the present to the
time of peak star formation.  For example, to estimate age for bright
ellipticals we use the mean $vz-yz$ and $bz-yz$ values from Schombert
\etal (1993).  The ellipticals studied in Schombert \etal (1993) ranged
from $vz-yz=0.580$ to 0.885 (varying as the color-magnitude relation where
fainter ellipticals were bluer) and this corresponds to expected $bz-yz$
values of 0.300 to 0.320.  The measured $bz-yz$ values were 0.356 to
0.364, which then corresponds to $\Delta(bz-yz)$ values from $-0.04$ to
$-0.06$ or $13\pm0.5$ Gyr (as shown in Figure 6).  This estimated age for
bright cluster ellipticals would coincide with the values determined by
spectral line studies (Trager \etal 2000) if corrected to the Salaris \&
Weiss cluster age system.

The range of $\Delta(bz-yz)$ values for Fornax dE's is shown as an inset
in Figure 6.  The mean $\Delta(bz-yz)$ for the dE's is $0.01\pm0.04$ which
corresponds to a mean age of 10 Gyrs with a range from 9 to 11 Gyrs.
While the accuracy on the $\Delta(bz-yz)$ index is low (since it is the
difference of two colors and carries the photometric error as well as the
error to the globular cluster calibration), the mean is statistically
different from bright ellipticals at the 99.9\% level.  The mean color of
Fornax dE's indicates, at the very least, that dwarf ellipticals are
younger than the oldest disk clusters in agreement with the age estimates
based on the Balmer lines of Virgo dE's (Bothun \& Mould 1988, Held \&
Mould 1994).

\subsection{Color-Magnitude Relation}

The relation between color and magnitude is well-known (see Bower, Lucey
\& Ellis 1992 and references therein) and is hypothesized to be a
relationship between the stellar mass of a galaxy (luminosity) and the
mean metallicity (color).  Alternative explanations to the color-magnitude
relation (CMR) that do not require metallicity effects have been proposed.
One of the more viable scenarios is that the CMR is due to variations in
age (Worthey 1994), but this hypothesis is not supported by Figure 7
(contingent on the various assumptions in the SSP models).  The
universality of CMR for the cluster environment has been demonstrated by
Bower, Lucey \& Ellis (1992) and, for rest of this paper, we will assume
that the CMR is due to metallicity changes in the underlying stellar
population.

Our best hypothesis for the origin of the CMR arises from the effects of
galactic winds (Larson 1974).  For a given unit of stellar mass, its mean
metallicity is determined by the degree of past gas consumption.  Since
ellipticals have exhausted their gas supply, their metallicities should
have identical values regardless of mass.  Thus, the existence of a
mass-metallicity relation (as reflected into the CMR) indicates another
that process has altered the chemical evolution of ellipticals.  Galactic
winds have the right direction and magnitude in the sense that a galactic
wind is delayed in more massive galaxies due to their deeper potential
wells (Yoshii \& Arimoto 1987, Tantalo \etal 1996).  This results in
stellar populations more enhanced in heavy elements and, therefore, redder
in color.  For low mass galaxies, galactic winds result in a mean
metallicity that is monotonic with mass.  At higher masses, the yield
flattens out at values around [Fe/H]=0.3 (Mould 1984) which agrees well
with the peak of the $vz-yz$,[Fe/H] relation in Figure 4.

\begin{figure}
\plotfiddle{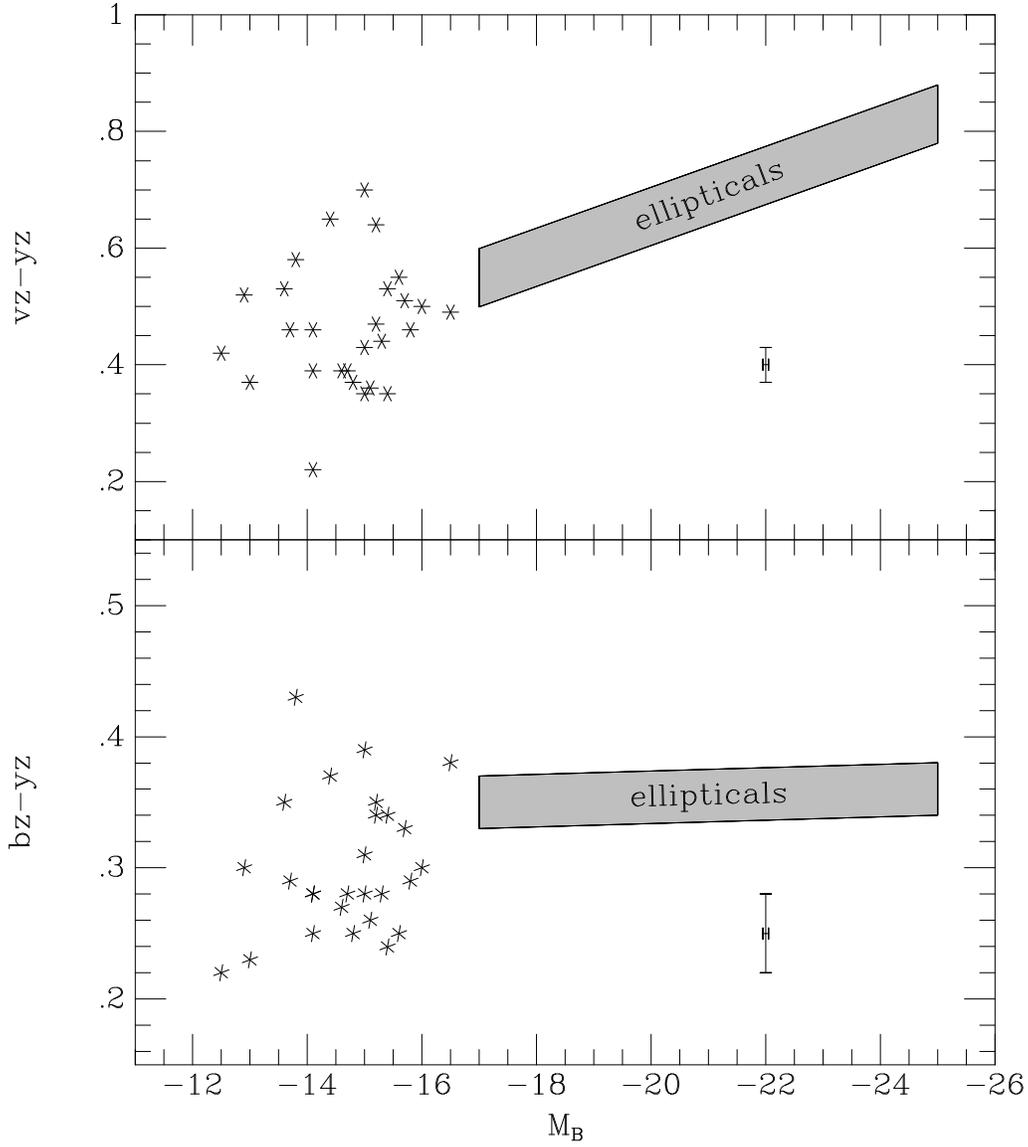}{12.0truecm}{0}{75}{75}{-220}{-60}
\caption{The color-magnitude relation for dwarf ellipticals.  The
shaded region is the CMR for bright ellipticals from Schombert \etal
(1993).  Typical error bars are indicated.  The Fornax dwarf ellipticals do
not appear to continue the tight CMR for more massive ellipticals in
agreement with other studies.}
\end{figure}

In our previous work on bright ellipticals (Schombert \etal 1993), the CMR
was determined for the $uz,vz,bz,yz$ system.  It is fairly strong for
$vz-yz$ and weak for $bz-yz$, as expected for a metallicity change.  Since
that study, we have extended the CMR to higher luminosities from studies
of distant clusters (Rakos \etal 2000).  The CMR for $vz-yz$ and $bz-yz$
are shown in Figure 11 along with the Fornax dE's.  Immediately obvious
from this diagram is that the dE's do not follow the same linear
relationship as bright ellipticals.  This is also seen in broadband colors
(Caldwell \& Bothun 1987, Prugniel \etal 1993) and it has been proposed
that the lack of correlation is due to the possibility that dwarf
ellipticals have an intermediate age population.  Interestingly, the
spread in $bz-yz$ is just as large, compared to the the bright elliptical
sequence, as it was for $vz-yz$.  Since $bz-yz$ is the most stable color
with respect to evolutionary effects (Rakos \& Schombert 1995), this hints
at real differences in the underlying stellar population between dwarfs
and more massive ellipticals, outside of an extrapolation of the
metallicity sequence.

\section{DISCUSSION}

\subsection{$\alpha$-element Enhancement}

In Schombert \etal (1993), the zeropoint to the $vz-yz$ colors was linked
to [Fe/H] through the Mg$_2$ index.  This was done in part because the
most direct measure of galaxy metallicity has traditionally been through
the Mg$_2$ index (Faber \etal 1995) as it is one of the clearest spectral
features in the optical portion of the spectrum due to the dominance of G
and K giant stars in galaxy stellar populations.  The correlation between
$vz-yz$ and Mg$_2$ was good, but a theoretical fit was required to obtain
[Fe/H] from Mg$_2$.  In our original study, we used the relationship from
Worthey (1994) with the assumption of a [Mg/Fe] ratio of 0.  However, the
resulting [Fe/H] values for the cores of bright ellipticals were, on
average, 0.3 dex too low compared to our estimates from their $J-K$ colors
($J-K$ measures the mean temperature of the RGB, also a metallicity
dependent feature in galaxies).  The values for brightest ellipticals were
also less than solar metallicities which, being more massive than the
Milky Way, was in conflict with our expectation of chemical evolution.  We
concluded, in that study, that the error was in the assumption that
[Mg/Fe] was zero and we predicted that there should be deficient in Fe
relative to Mg for our elliptical sample.

This deficient between Fe and lighter elements, usually expressed as the
[Mg/Fe] or [$\alpha$/Fe] ratio, is known as the $\alpha$ element
enhancement since the elements in question are all even atomic numbers
synthesized by $\alpha$-capture (see Worthey 1998).  The problem with
$\alpha$ element abundance arises from the different methods in which
light and heavy metal production occurs in a galaxy.  Light $\alpha$
elements, such as Mg, and heavy elements, such as Fe, are both produced in
Type II supernovae (SN), the carbon detonation of a massive stellar core.
Light and heavy elements are also produced by Type Ia SN, the explosions
due to accreting white dwarfs, but Type Ia SN overproduce Fe relative to
$\alpha$ elements in these events (from 2 to 10 times more Fe than Type II
SN).  Massive stars evolve quickly and, thus, Type II SN have very short
lifetimes on the order of a few to 100 million years (Kobayashi \etal
1998).  Type Ia SN, on the other hand, require the companion to complete
its slow evolution to the RGB and are estimated to have lifetimes from 0.6
Gyrs to a few Gyrs (Ishimaru \& Arimoto 1997).

The different yield of Fe to lighter elements and the different timescales
for Type Ia and Type II SN's allows for a range in [$\alpha$/Fe] depending
on their relative contributions to the galactic ISM.  For example, the
difference in timescales between Type Ia and II SN would be evident in
galaxies with initial episodes of star formation which take less than a
Gyr such that they will have chemical yields deficient in Fe compared to
lighter elements (note that the ratio [$\alpha$/Fe] is governed by the
production, or lack thereof, of Fe not $\alpha$ elements).  Star formation
which is delayed over several Gyrs will allow time for Type Ia SN to
inject additional Fe into the ISM for later generations of stars.  In our
own Galaxy, a shift in the [Mg/Fe] ratio is detected in metal-poor versus
metal-rich field stars in the sense that older stars have enhanced [Mg/Fe]
which slowly decreases to solar values as the stellar population becomes
younger (Wheeler, Sneden \& Truran 1989), presumably reflecting the
delayed enrichment of the Galactic ISM by Type Ia SN.

The observational trend of [$\alpha$/Fe] is difficult determine since it
relies on high S/N data of weak Fe lines (such as Fe 5270 and Fe 5335).
Spectroscopy by Brodie \& Huchra (1991) clearly shows that bright
ellipticals have Mg$_2$ indices higher than expected from extrapolation of
globular cluster values and Worthey, Faber \& Gonzalez (1992) demonstrate
that the strength of the Fe 5270 and 5335 lines are weaker than expected
for the Mg b line regardless of age or mean metallicity of the underlying
stellar population.  Fortunately, the direct evidence for a varying
[Mg/Fe] ratio within ellipticals has improved dramatical in the last year.
Both Trager \etal (2000) and Kuntschner \etal (2000) present Mg and Fe
relations for over 80 field and cluster ellipticals that demonstrates an
increasing [Mg/Fe] ratio with galaxy mass (see their Figures 6 and 12
respectively).  The relationship appears independent of cluster
environment and ranges from [Mg/Fe]=0.3 for bright ellipticals ($M_B=-22$)
to [Mg/Fe]=0.1 for faint ellipticals ($M_B=-18$). 

To investigate the run of [Mg/Fe] between globular clusters, Fornax dwarfs
and bright ellipticals, we have compared our $vz-yz$ values to Mg$_2$
values in the literature.  The philosophy assumed here is that $vz-yz$ is
coupled to the total metallicity of a stellar population ($Z$) through
the behavior of the turnoff point and RGB colors, rather than weighted
towards the $\alpha$ elements.  This is not strictly true since isochrone
shapes will change with varying [Mg/Fe] (Salaris \& Weiss 1998), although
the work of Tantalo, Chiosi \& Bressan (1998) concluded that change in
[Mg/Fe] lead to minor changes in the CMD of a SSP.  Our analysis is
further complicated by the practical consequence of calibrating $vz-yz$ to
[Fe/H] rather than the total metallicity, $Z$, which actually determines
color of the underlying stellar population.  In order to investigate the
magnitude of this problem, we have turned to the nonsolar abundance models
of Weiss, Peletier \& Matteucci (1995).  These models generate isochrones
for [Mg/Fe] values of 0.0 and 0.45 over a range of metallicities, helium
abundance and mixing lengths.  For our needs, it is sufficient to examine
the change in $vz-yz$ color for varying [Mg/Fe] while holding all other
variables, particularly [Fe/H], constant.  When this experiment is
performed, we find that $vz-yz$ varys by less than 0.01.  This is a
negligible amount corresponding to only a change of 0.03 dex in [Fe/H] (see
Figure 7) or 0.1 Gyrs in age.  We note that recent work by Salasnich \etal
(2000) finds that the turnoff of a 10 Gyr population will shift redward by
log $T_{eff}$=0.011 for a change in [Mg/Fe] of 0.4.  This would correspond
to a shift in $vz-yz$ of 0.04 or 0.1 dex in [Fe/H].  While this shift is
larger then previous work, this is still small in comparison to other
sources of error in our calibrations.

It is not too surprising to find that small changes in [Mg/Fe] have minor
consequences to the integrated $vz-yz$ color.  The multi-metallicity model
in Figure 7 assumes no color changes due to [Mg/Fe] changes, yet
produces an excellent match to the $vz-yz$ versus [Fe/H] relation for
bright ellipticals despite a change from 0.1 to 0.3 in [Mg/Fe] (Trager
\etal 2000, Kuntschner \etal 2000).  In addition, the [Fe/H]-(vz-yz)
relation for ellipticals can not be due to [$\alpha$/Fe] effects since the
$vz-yz$ colors are too blue for their [Fe/H] values compared to SSP models
or an extrapolation of the globular sequence.  If $\alpha$ elements
dominate the isochrones of the stellar population, then the colors of the
brightest ellipticals (with the highest [$\alpha$/Fe] values) would have
colors redder than the extrapolation of the globular cluster sequence,
which is clearly not the case in Figure 7.  

\begin{figure}
\plotfiddle{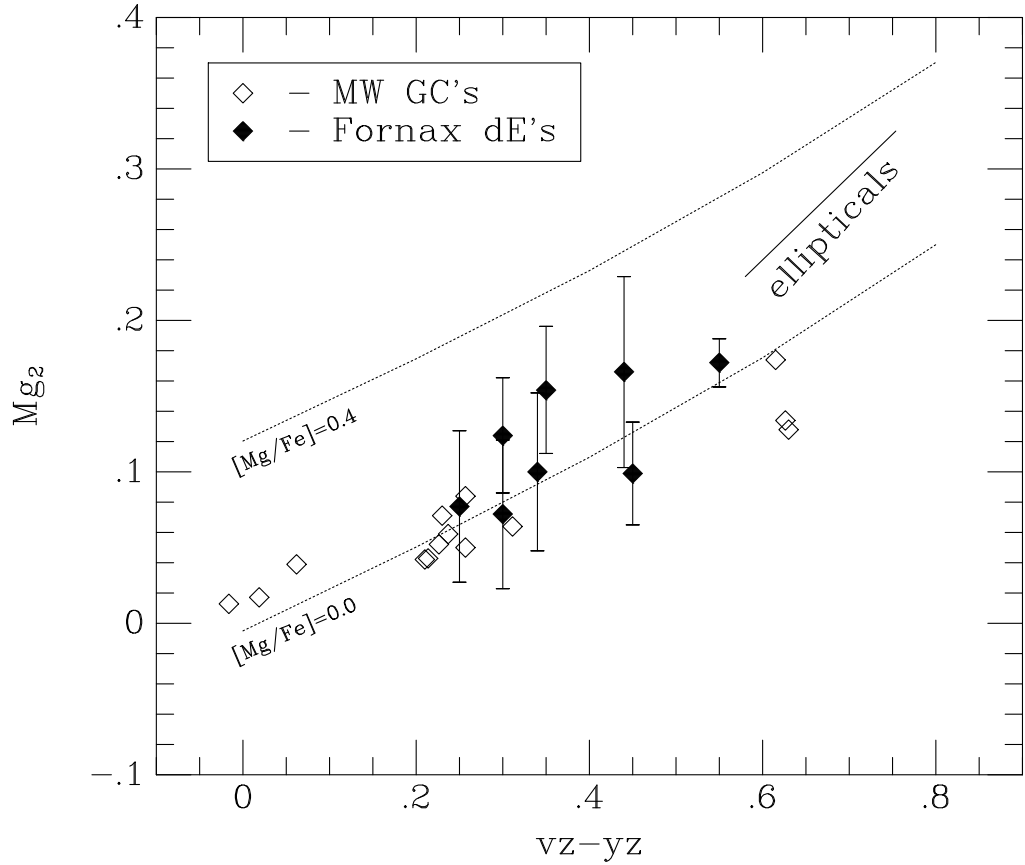}{8.0truecm}{0}{85}{85}{-230}{-210}
\caption{The Mg$_2$ index versus $vz-yz$ color.  Globular clusters (open diamonds,
Mg$_2$ indices are from Brodie \& Huchra 1991) and Fornax dE's (filled
diamonds, Held \& Mould 1994) are shown The dashed line is the
relationship between Mg$_2$ and $vz-yz$ for bright ellipticals from
Schombert \etal (1993).  Dotted lines are the curves of constant [Mg/Fe]
from  Weiss, Peletier \& Matteucci (1995).  While the dwarf elliptical
data indicates slightly higher than solar [Mg/Fe], the bright ellipticals
range from 0.1 to 0.3 dex.  Globular clusters display a trend of
decreasing [Mg/Fe] with metallicity, in agreement with their ages.}
\end{figure}

The comparison between $vz-yz$ and Mg$_2$ for globular clusters, Fornax
dwarfs and bright ellipticals is found in Figure 12.  The solid line
represents the $vz-yz$, Mg$_2$ relation for ellipticals brighter than
$M_B=-18$ adopted from Trager \etal (2000) and Kuntschner \etal (2000).
The Mg$_2$ values for globular clusters and the Fornax dwarfs with $vz-yz$
measurements are taken from Brodie \& Huchra (1991).  Also shown are
curves of constant [Mg/Fe] taken from Weiss, Peletier \& Matteucci (1995)
with zeropoints set such that the top and bottom of the elliptical
sequence corresponds to 0.3 and 0.1 respectively (Trager \etal 2000).

It is immediately clear that the dwarf ellipticals and globular clusters
occupy the same region of Mg$_2$, $vz-yz$ space, although the scatter is
high.  The globulars follow a trend of decreasing [Mg/Fe] with increasing
$vz-yz$ (i.e. [Fe/H]) as is expected from specific measurements of Galactic
halo stars (Wheeler Sneden \& Truran 1989).  For globular clusters in
Figure 12, the trend of increasing $vz-yz$ is also a trend of increasing
age.  Old clusters are enhanced in $\alpha$ elements due to the lack of
enrichment by Type Ia SN.  Younger clusters, having formed several Gyrs
after the oldest globulars, have [$\alpha$/Fe] values approaching solar.

The dwarf ellipticals lie in a region of the diagram that indicates they
are an extension of the bright elliptical sequence to fainter magnitudes.
Their position also indicates [$\alpha$/Fe] values near, or slightly
above, solar.  This result is only preliminary and should be confirmed
with direct measurement of Fe lines.  Currently, the data in Figure 12 is
consistent with the trend determined by Trager \etal (2000) of decreasing
[$\alpha$/Fe] with decreasing galaxy mass.  We also note that, if
confirmed, Figure 12 allows a crude measure of [$\alpha$/Fe] in galaxies
which may be difficult to directly measuring weak Fe lines due to their
intrinsic faintness or low surface brightness.  For example, Mg$_2$ and
$vz-yz$ would be easier to obtain in distant galaxies or LSB objects using
only low resolution spectroscopy.

The origin of the trends of [$\alpha$/Fe] in Figure 12 is obscure.
Clearly, the different yields of $\alpha$ elements and Fe from Type Ia and
II SN are the primary source for the variations with galaxy mass, unless
some process can selectively remove light or heavy elements from the ISM.
There are several different scenarios to produce the trend of
[$\alpha$/Fe] with galaxy mass (see Trager \etal 2000 for a fuller
discussion).  We will consider three possible scenarios in conjunction
with the impact on the star formation history in dwarf ellipticals; 1)
varying lengths in star formation duration, 2) different galactic winds
strengths and 3) variations in the initial IMF.  The data presented in
this section is insufficient to choose between these scenarios.

The most obvious explanation for the low [$\alpha$/Fe] in dwarfs is that
low mass galaxies have longer periods of initial star formation, compared
to high mass ellipticals.  A longer initial burst of star formation leads
to a later epoch before the onset of galactic winds and, therefore, more
time for Type Ia SN detonation and the increase in Fe abundance (see
Worthey, Faber \& Gonzalez 1992).  However, the total elemental production
is increased with longer durations of star formation, in contradiction
with the color-magnitude relation, unless the strength of the initial
burst also decreases, i.e. fewer generations per unit of time (see
scenarios 1 and 2 of Trager \etal 2000).  This make this hypothesis
problematic with regard to the mass-metallicity relation (see \S 3.2).

While star formation, and SN production, is the known method to produce
heavy metals, a mechanism to halt star formation is required to produce
the mass-metallicity relation in galaxies.  Otherwise, closed box models
required that all galaxies have identical (and much great than solar) mean
metallicities.  The most plausible method to produce the mass-metallicity
relation is to invoke galactic winds to halt star formation by removal of
the ISM (see \S 3.5).  Certainly, the production of SN during initial star
formation provides the energy source for the heating of the ISM into a
wind state.  Higher mass galaxies would maintain deeper gravitational
wells which, in turn, requires more energy for a wind.  More energy
results in greater metal injection and higher mean metallicities for each
generation of stars.  With respect to [$\alpha$/Fe] enhancement, the
timing to the onset of a galactic wind and/or its strength can influence
the ratio of retained $\alpha$ elements.  For example, if the onset of a
galactic wind were delayed, then the mean [$\alpha$/Fe] value would
decrease as more Type Ia SN contribute to the ISM.  Another possibility is
if the wind varys in strength with galaxy mass.  Then, low mass galaxies
would have weak winds that slow the depletion process allowing late
enrichment by Type Ia SN and, thus, lowering the mean [$\alpha$/Fe].

Lastly, since $\alpha$ elements arise primary from massive stars, a
varying IMF can result in different [$\alpha$/Fe] values.  In this
scenario, high mass galaxies are richer in massive stars during their
early stages of star formation.  This leads to a higher contribution from
Type II SN and an increase in [$\alpha$/Fe] with galaxy mass (assuming
that the time before galactic winds is longer than the timescale of Type
Ia SN).  This idea is not without merit since higher mass galaxies have a
higher number density of clouds and more violent cloud collisions.  This
type of star formation has been detected in galaxy mergers along with
indications of a higher number of high mass stars (Elmegreen \etal 2000).
Additionally, an early enhancement of the IMF will not be detected in
present day galaxy colors since, if the changes involve only the highest
mass stars, then those stars have long since evolved out of the current
visible stellar population.  

\begin{figure}
\plotfiddle{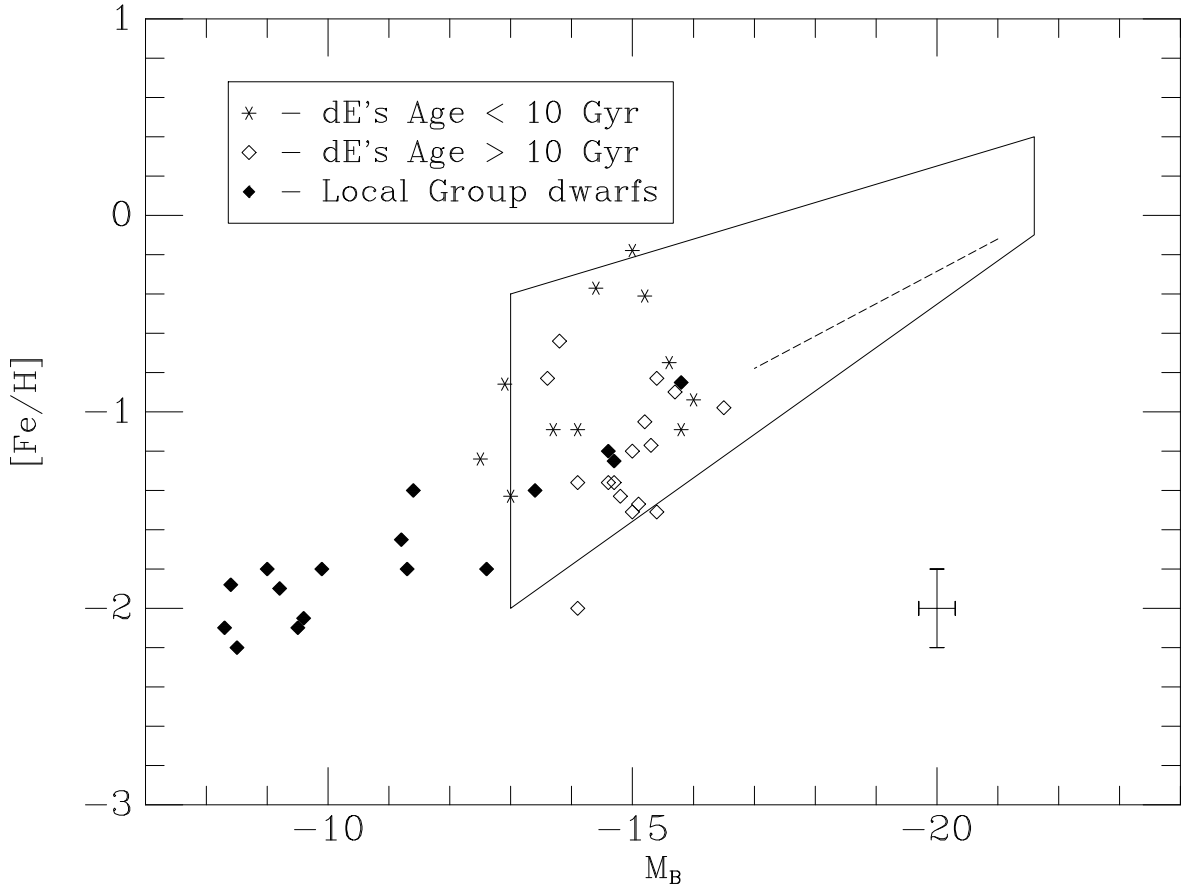}{10.0truecm}{-90}{65}{65}{-270}{390}
\caption{The mass-metallicity diagram for normal and dwarf
ellipticals.  The outlined region is the mass-metallicity relation from
Brodie \& Huchra (1991) (their Figure 4).  The dashed line is the
color-magnitude relation from Schombert \etal (1993).  Fornax dwarf
ellipticals with ages greater than 10 Gyrs (open diamonds) are compared
with Fornax dE's with ages less than 10 Gyrs (crosses).  Local Group
dwarfs (filled diamonds) from da Costa (1998) are also displayed.  As the
mass-metallicity relation broadens for low luminosity ellipticals, there
is a tendency for younger ellipticals to lie at the top of the sequence
and the older galaxies to form the lower sequence.}
\end{figure}

\subsection{Mass-Metallicity Relation}

The key to understanding the chemical evolution of ellipticals is the
relationship between galaxy mass and the mean metallicity of its stellar
population.  The CMR is just one manifestation of this effect and other
correlations, such as with central velocity dispersion and metallicity
lines or the $Z$ hyperplane (Trager \etal 2000), are also present in
ellipticals.  To examine the mass-metallicity relation for dwarf galaxies,
we have plotted [Fe/H] versus absolute blue magnitude ($M_B$) for Local
Group dwarfs, Fornax dwarfs and bright ellipticals in Figure 13.  We have
made two assumptions in the following interpretation of this diagram: 1)
that [Fe/H] traces the total metallicity of a galaxy ($Z$) and 2) that the
absolute blue magnitude traces the mass.  The first assumption is
supported by numerous line indices and gradient studies.  The second
assumption means that the $M/L_B$'s are relatively constant between dwarfs
and bright ellipticals.  Bright ellipticals range in $M/L_B$ from 4 to 10
(Kronawitter \etal 2000), but dwarf ellipticals are less well known.
Peterson \& Caldwell (1993) find at range of $M/L_B$ from 1 to 7 for dE's
in Fornax and Virgo.  Population models estimate $M/L_B$ in the range of 4
to 6 (Worthey 1994).  We conclude that, within 20\%, the blue magnitude
traces the mass from dwarf to bright elliptical.

The shaded zone in Figure 13 represents the region of magnitude and
metallicity outlined by the Brodie \& Huchra sample (their Figure 3,
adjusted to an $H_o=75$).  While displaying the known trend of decreasing
[Fe/H] with mass, the region is notable for the increase in its
metallicity range with lower elliptical luminosities.  The dashed line
represents the Schombert \etal (1993) CMR calibrated to [Fe/H] following
the prescription in \S2.3.  Since our bright elliptical and distant
cluster datasets do not sample below $M_B=-18$, we can not test the
validity of the increase in the range of [Fe/H] for low luminosity
ellipticals except to note that other studies have confirmed this increase
beyond what is expected for observation uncertainty.  The data for Local
Group dwarfs (Da Costa 1998) is also shown in Figure 13.

By themselves, the Fornax dwarf data do not follow any mass-metallicity
relation, as we would have expected from the scatter in the $vz-yz$ CMR
from \S 2.3.  Its interesting to note that there is a group of Fornax dE's
that follow the relationship defined by Local Group dwarfs and the CMR for
bright ellipticals and a group that lies above the CMR with a distinct gap
between them.  This suggests that the spread in mass-metallicity may be
due to a galaxy second parameter effect, such as a dependence on the
central velocity dispersion (Terlevich \etal 1981) or a range of star
formation efficiency, where galaxies along the top of the sequence were
highly efficient at converting gas into stars versus those galaxies along
the bottom edge of the sequence.  Another possible factor in the spread of
metallicity would be some event, such as a merger or tidal collision,
which removed gas and/or halted star formation.

Lacking dynamical information on our dwarf sample, we instead test the age
of the Fornax dwarfs using the $\Delta(bz-yz)$ index.  The sample is
divided into those dwarfs with mean ages greater than 10 Gyrs and those
less than 10 Gyrs, shown as different symbols in Figure 13.  While there
is no trend of [Fe/H] and age, it is clear that the younger dwarfs tend
lie above the older dwarfs in the mass-metallicity diagram.  Notice that
the trend is opposite to what is expected from observational error in that
the younger galaxies per luminosity bin have high [Fe/H] values (i.e.
redder colors).  Note also that younger galaxies are not necessarily bluer
than the mean, but merely bluer than the other galaxies in their mass bin.
This confirms what was noted by Bothun \& Mould (1988) and Held \& Mould
(1994) that there is a range of Balmer line strengths in dE's at a fixed
metallicity compared to globular clusters.  

This trend between galaxy mean metallicity and age is understandable in
terms of a simple model of chemical evolution.  If cluster galaxies all
form at the same epoch, then younger mean age implies a longer time scale
for star formation and more time to recycle the ISM before the onset of
galactic winds.  Thus, per mass bin, one would expect the younger galaxies
to have higher metallicities as younger age represents more time for the
yield from SN to increase the [Fe/H] for each stellar generation.  We conclude
that the mass-metallicity relation is distorted for dwarf ellipticals, not
due to range of ages as it directly influences the CMR, but rather by the
younger age galaxies having higher metallicities due to a longer history
of chemical evolution.

\subsection{Mean Age of Dwarf Ellipticals}

In a recent study of 50 local field and cluster ellipticals, Trager \etal
(2000) conclude that there is range of ages in field ellipticals (from 1.5
to 18 Gyrs), but that cluster ellipticals tended to be old with a mean age
between 10 and 13 Gyrs.  Kuntschner (2000) confirms the coeval nature of
the cluster ellipticals in Fornax with a mean age of 13 Gyrs using the
H$\beta$ versus Fe diagrams pioneered by Worthey (1994).  Bressan, Chiosi
\& Tantalo (1996), using a similar analysis, find that there must be age
variations, in addition to metallicity changes, to explain the
distribution of elliptical's 1550\AA\ colors, H$\beta$ and Fe indices, but
that all ellipticals are older than 6 Gyrs.  Certainly, in this study, the
CMR of cluster ellipticals is well modeled by a luminosity weighted metal
distribution (see \S 2.3) and the scatter in the CMR restricts the age
spread to less than 2 Gyrs for cluster ellipticals.  In addition, studies
that follow the color evolution of cluster ellipticals are well matched to
models of passive evolution of their stellar populations to a redshift of
formation of $z_g=5$ (Rakos \& Schombert 1995).  

As mentioned in the introduction, the line indices methods have the
greatest resolution of the age-metallicity degeneracy, but also carry the
light element enhancement assumptions.  The [$\alpha$/Fe] values used by
Trager \etal (2000) all were empirically determined from their own indices
and those values are confirmed in \S 2.3.  However, an alternative to the
line indices method of dating ellipticals is to examine the mean colors
which measure the average position of the turnoff point (the change in the
position of the RGB is relatively small for small changes in age).  To
this end, we have used the $\Delta(bz-yz)$ index as a direct measure of
the mean age of bright and dwarf ellipticals through the color of the
turnoff stars.  To repeat the discussion in \S 2.1, the $vz-yz$ index is
less sensitive to age than to metallicity when compared to $bz-yz$.  We
can, in effective, predict the $bz-yz$ color based on the $vz-yz$ index.
In terms of the underlying stellar population, this is equivalent of
examining deviations of the turnoff point for a fixed metallicity.
Negative $bz-yz$ residuals signal an older population, positive residuals
signal a younger one.  We have calibrated the $\Delta(bz-yz)$ index using
globular clusters and SSP models; however, we caution the reader that
ellipticals will have the broad turnoff points that one would expect from
a burst of star formation extended in time (globular clusters form all
their stars in very short timescales producing sharp CMD's).  We note, for
example, the color-magnitude diagram of the Fornax dwarf spheroidal
(Buonanno \etal 1999) has a wide RGB and main sequence indicating an
onset of star formation 12 Gyrs ago with continuous bursts up to 0.5 Gyrs
ago.  

In \S 2.2, it was demonstrated that the $\Delta(bz-yz)$ index implies that
dwarf ellipticals are about 3 Gyrs younger than bright ellipticals in mean
age.  There is no evidence from their narrow band colors that cluster
ellipticals have a range in age and the CMR suggests they are coeval with
a mean age of 13 Gyrs.  The Fornax dwarfs, on the other hand, range in age
from about 9 to 11 Gyrs.  There is a great deal of scatter in these values
due to observational errors and the inherent uncertainties contained in
our method, but the mean age for dwarfs is clearly much less than bright
ellipticals.  If the Fornax dwarfs' younger mean age is universally
applicable to all dE's, then this would explain the spectroscopic
observations of Virgo dwarfs by Bothun \& Mould (1988).  In that study, it
was found that the Ca II K ratio was similar to metal-rich clusters such
as 47 Tuc.  However, the Balmer line strength was slightly stronger,
indicating a younger age, younger than globular cluster ages.  If we take
the age of 47 Tuc to be 9.2 Gyrs (Salaris and Weiss 1998), then a mean age
for Fornax dE's of 9 to 10 Gyrs would be in agreement with the
spectroscopic data of Virgo dE's.  In addition, both Thuan (1985) and
Caldwell \& Bothun (1987) found that the $B-H$ colors for dE's in Virgo
are too blue relative to their $J-K$ index, again a signature of younger
mean age.  

One possible complication to our dating method is that it is possible that
our colors may be reflecting a recent burst of star formation on top of an
old stellar population.  To test for recent star formation, we added 1\%
of a 1 Gyr population to a low metallicity 13 Gyr population (so-called
frosting models, Trager \etal 2000).  The resulting colors would produce a
1 Gyr shift in the mean age as calculated by the $\Delta(bz-yz)$ index.
Thus, the addition of 3 to 4\% of a 1 Gyr population would be required to
produce the mean ages observed in Figure 6.  This would seem excessive
given the lack of H\,I detection in dE's.  The dwarf ellipticals in the
M81 group, similar in mass to the Fornax dwarfs, display no evidence of a
young population (less than 2 Gyrs) from HST CMD diagrams (Caldwell \etal
1998).  In addition, vacuum UV imaging places severe constraints in the
number of OB stars in dwarf ellipticals (less than 1\% contribution by a
hot component, O'Neil \etal 1996) and, thus, we rule out recent star
formation for dE's as a source of the bluer $bz-yz$ colors.  

\subsection{Star Formation Duration}

It is difficult to determine from the present data, or from model fits,
whether the younger mean age for in any galaxy is due to a later epoch of
initial star formation (compared to bright ellipticals) or due to the
effects of an extended duration of star formation.  We can attempt to
infer the duration of star formation based on the trend of [Mg/Fe] in \S
3.1, but only if we assume that the change in $\alpha$ elements is due
solely to the delay between Type Ia and Type II SN.  This section will
focus on duration effects and other processes, such as galactic winds,
will be considered in the next section.

To first order, a lower [Mg/Fe] value implies longer star formation
timescales in order for Type Ia SN enrichment to occur.  The change in
[Mg/Fe], suggested in Figure 12, implies that bright ellipticals ($M_B >
-18$) have a duration of initial star formation that is inversely
proportional to their mass.  The duration of star formation for low mass
ellipticals was long enough (a few Gyrs) for Type Ia SN to contribute to
the mean metallicity with corresponding lower [Mg/Fe] values.  Conversely,
the duration of star formation for high mass ellipticals must have been
shorter resulting in higher [Mg/Fe] values.  Under this interpretation,
the Fornax dwarf ellipticals must have even longer duration times, compared
to bright ellipticals, since there [Mg/Fe] values are close to solar.  

In addition to the [Mg/Fe] information, we also have the age estimates
from Figure 6 which indicates a mean age for dwarfs ellipticals of 10
Gyrs, about 3 Gyrs younger than bright ellipticals.  This is consistent
with the hypothesis of a longer duration of star formation compared to
bright ellipticals as the younger mean age would then be reflecting a
later epoch for the cession of star formation.  Given the division by age
in the mass-metallicity relation (Figure 13), the star formation history
of dwarf ellipticals may break into two sequences, an upper track with a
longer duration (younger mean age) and a lower track with a shorter
duration and colors that indicate an older population.  This would predict
that upper sequence dwarfs have enhanced [Mg/Fe] compared to upper
sequence, but there is no data at this time to test this hypothesis.

The main difficulty in the interpretation of the [Mg/Fe] and mean age
data, in terms of the duration of the initial burst of star formation, is
the fact that longer timescales of star formation should imply higher
metallicities.  This was pointed out in Trager \etal (2000) that the trend
of increasing [$\alpha$/Fe] is at odds with the mass-metallicity relation
for ellipticals.  A shorter duration of star formation for high mass
ellipticals should result in less chemical evolution and, therefore, lower
metallicities.  Clearly, another mechanism is involved in the evolution of
galaxy stellar populations, such as galactic winds.

\subsection{Galactic Winds}

Massive star formation ultimately leads to the injection of energy into
the ISM through SN and stellar winds.  If the energy input exceeds the
gravitational binding energy of the galaxy, then a galactic wind develops
removing the remaining ISM (Larson 1974).  Evidence for galactic winds is
mostly indirect.  They are needed to remove gas from ellipticals in a
quick in timely fashion so that most present-day ellipticals are free of
neutral gas (although they have large quantities of hot, X-ray emitting
gas, presumingly the remnant of baryonic blowout).  Since cluster
elliptical colors indicate coeval star formation epochs with a sharp
cutoff well over 10 Gyrs ago (Rakos \& Schombert 1995), galactic winds are
the preferred method to cease star formation in a uniform fashion to
cluster ellipticals as a class of objects.  Galactic winds are also
convenient for explaining the high heavy metal abundances of the
intracluster medium (Arnaud \etal 1992).  Ellipticals, again, are a likely
source of this material since they are a major mass component of rich
clusters of galaxies.

The mass-metallicity relation also requires some mechanism like galactic
winds to explain the dependence of mean metallicity on the total mass of a
galaxy.  Regardless of mass, the mean metallicity of a region is governed
by gas consumption.  If star formation proceeds in ellipticals such that
all the gas is converted into stars, then they should have high, and
nearly constant, mean metallicities (i.e. closed box model).  Clearly,
cluster ellipticals are not of uniform metallicity, as demonstrated by the
CMR and spectral line studies.  The onset of a galactic wind, as a
function of mass, produces the proper correlation with the reasoning that
high mass galaxies have a higher gravitational potential which retains the
gas for a longer amount of time allowing for higher enrichment.  In this
scenario, known as the classic wind model, the key parameter that
determines the mean metallicity of a galaxy is the timescale before the
onset of a galactic wind where longer timescales result in higher
metallicities.

On the other hand, an early galactic wind, after short duration of star
formation, may result in high metallicities if star formation is more
efficient with higher galactic mass.  This is opposite to classic wind
scenario, called the inverse wind model (Matteucci 1994).  In the inverse
wind model, galaxy formation follows a different course than the assumed
homogeneous collapse of a giant protogalactic gas cloud.  Some process,
for example mergers, produces higher densities and higher cloud collision
rates than those resulting from pure dynamical timescales.  This results
in more massive galaxies developing an early galactic wind, but with a
higher rate of star formation and corresponding higher rates of chemical
evolution.

The key parameter in distinguishing between the classic and inverse wind
model is the time between initial star formation and the onset of galactic
winds, $t_{gw}$.  For ellipticals, this is assumed to be equivalent to the
duration of star formation since the galactic winds remove the ISM and
halt subsequent star formation.  Classic wind models assume that $t_{gw}$
decreases with decreasing galactic mass.  For example, Kodama \& Arimoto
(1997) reproduce the $V-K$ CMR using $t_{gw} = 0.5$ Gyrs for the bright
end, dropping to 0.1 Gyrs for the faint end.  The inverse wind model
(Matteucci 1994) reverses the trend with a SF duration of 0.5 Gyrs at the
faint end dropping to 0.1 Gyrs for galaxies with $M = 10^{12} M_{\sun}$.
We note that the classic wind model is at odds with the observation that
[$\alpha$/Fe] decreases with galactic mass such that cession of star
formation in 0.1 Gyrs for faint ellipticals will not produce the
[$\alpha$/Fe] values near 0.1 seen by Trager \etal (2000) and Kuntschner
(2000).

Some indication of the relative importance of star formation duration and
onset of galactic winds is provide by the range of age for dwarf
ellipticals in Figure 13, the mass-metallicity diagram.  The dwarfs have a
range of metallicities (from [Fe/H]=$-$1.6 to $-$0.4).  A pure galactic
wind model would predict early winds produce low metallicity dwarfs and
the onset of an early wind would also produce an older stellar population.
Indeed, that is what is observed in Figure 13, the oldest dwarfs have the
lowest metallicities, while the youngest dwarfs have the highest
metallicities.  A pure wind interpretation would also imply a similar rate
of star formation (per unit mass) with a later epoch of winds providing
the cutoff for mean age (younger for late winds) and a longer time for
chemical evolution (higher [Fe/H]).  

The problem with this simple wind model is that one would expect a
correlation between dwarf mass and metallicity.  An early wind implies a
low gravitational potential and, thus, a lower mass dwarf compared to
those which produced late winds.  Therefore, we would expect to find a
tight mass-metallicity correlation for dwarf ellipticals, which is not the
case in Figure 13.  Since neither star formation duration nor galactic
winds are able to satisfactory explain the characteristics of dwarf
ellipticals, we are led to speculate that external factor, such as the
cluster environment, may have a controlling factor in the early star
formation history of cluster dwarfs.

\subsection{Cluster Environment}

If bottom-up scenarios for galaxy formation are correct, then dwarf
galaxies should exist as fossil remnants of the epoch of galaxy formation.
Their metallicities would than provide the most direct look into the
chemical history of a cluster, with the assumption that their current
positions map into their dynamical past.  Figure 14 displays the $vz-yz$
colors of the dwarfs as a function of radius from the core of the cluster
(assumed to be the position of NGC 1399).  There is a clear gradient in
metallicity color from the core to the halo which corresponds to about 0.7
dex.  Core metallicities are lower than halo metallicities (opposite to
our own Galaxy) with a mean of [Fe/H]=$-$1.3 which then rises to a mean of
[Fe/H]=$-$0.6 in the outskirts of the cluster.

\begin{figure}
\plotfiddle{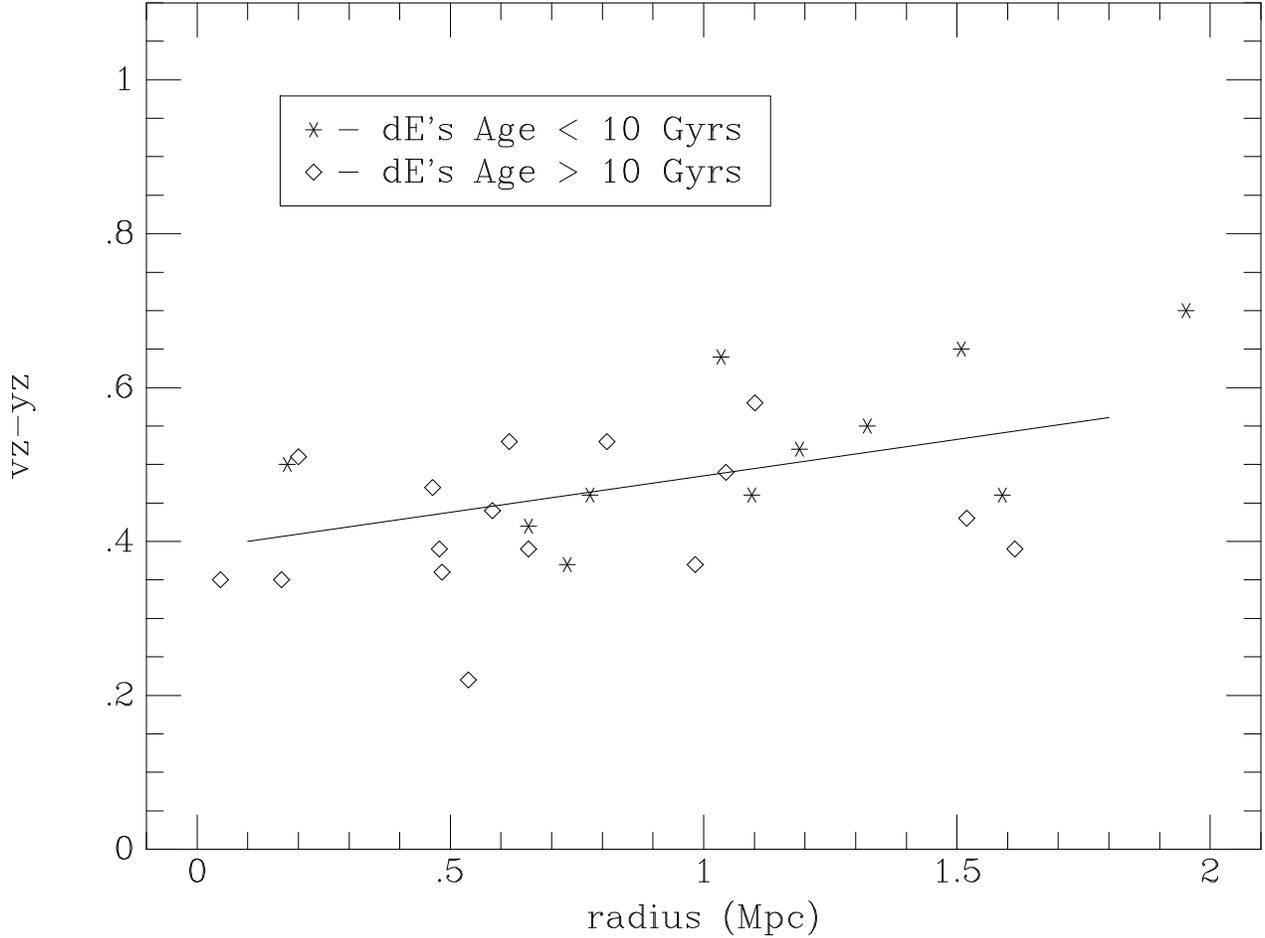}{10.0truecm}{-90}{70}{70}{-270}{390}
\caption{Fornax dwarfs $vz-yz$ color as a function of cluster radius.  The Fornax
dwarf ellipticals display a noticeable increase in metallicity with
increasing cluster radius.  This is opposite to the metallicity trend for the
intracluster gas, which decreases with radius, and may indicate that
dwarfs are a major source of metals to the intracluster gas.  We also find
that there is a weak correlation with galaxy mean age as a function of
radius such that older dwarfs inhabit the cluster core and younger dwarfs
are found in the halo.  This observation is consistent with the idea that
the intracluster medium halts star formation and enrichment in dwarf
galaxies through a ram pressure stripping mechanism.}
\end{figure}

This decrease in metallicity with decreasing radius is opposite to the
expectations due to metallicity observations of the hot x-ray gas in
clusters.  ASCA measurements of the Fe K line for several rich clusters of
galaxies indicates a drop from 1/2 solar at the core to 1/3 solar in the
outskirts (Dupke \& White 2000).  In addition, the abundance ratios in the
core intracluster gas suggests a strong Type Ia SN component (Dupke \&
White 2000).  Since metals in the intracluster gas have two possible
origins, galactic winds or ram pressure stripping (Gunn \& Gott 1972), the
abundance ratio  would rule out an origin for the gas strictly from matter
ejected by early winds, which is primarily Type II SN enriched material.
Later winds or stripping of Type Ia SN gas from bright ellipticals is
required to balance the abundance ratios.  We note, however, that this
[$\alpha$/Fe] ratio matches the expected ratios in dwarf ellipticals
and, thus, they may contribute a large fraction of the intracluster gas.

If dwarfs are major contributors to the intracluster metals then there may
be a feedback process involved such that the cluster environment is
responsible for the dwarf metallicity values.  For example, the low [Fe/H]
values of core dwarf ellipticals may be due to the fact that they have
injected most of their metals to the intracluster gas rather than
recycling the material for their own stars.  The opposite would then be
true for dwarf ellipticals in the outer cluster regions, their
metallicities are enhanced relative to the core because they have held
onto their metal enriched gas for later generations.  Thus, the
intracluster gas was deprived of their metals and have lower [Fe/H] values
in the halo, resulting in a decreasing ICM abundance gradient with cluster
radius.  The mechanism for this process is, presumingly, ram pressure gas
stripping by the intracluster medium and this process should be more
effective on low mass objects, such as dwarf ellipticals, than high mass
galaxies due to their shallower gravitational potentials.  

An addition test to the clusters influence is to examine the mean ages of
the dE's as a function of cluster radius.  If the outer dwarfs have higher
metallicities due to a delay in stripping by the cluster ICM, then we
would expect to find a younger mean age reflecting the longer duration of
star formation and continued metal processing.  Likewise, for the inner
galaxies, we would expect to find an older mean age reflecting the earlier
shutdown of star formation by the ICM.  In fact, the distribution of ages
indicates that this is the case.  Figure 14 plots dE's with greater than
or less than 10 Gyr ages based on the $\Delta(bz-yz)$ index.  As expected,
a majority of the dE's within a radius of 0.7 Mpc are older than 10 Gyrs.
This would also explain why Trager \etal (2000) find a range of elliptical
ages in the field, but Kuntschner (2000) find only old, coeval ellipticals
in the Fornax cluster.  The cluster environment appears to play a major
role in the cession of initial star formation in ellipticals much like the
cluster environment effects infalling spirals at intermediate redshifts
(i.e. the Butcher-Oemler effect).

\section{CONCLUSIONS}

The first part of this paper was spend examining the behavior of our
narrow band filter system under changes in age and metallicity using
globular cluster data and SSP models.  The calibrations and
multi-metallicity population models produce excellent fits to the past
data on both bright and dwarf ellipticals.  Using these calibrations, we
present a new set of narrow band photometry for Fornax dwarf ellipticals.
We can summarize our results as the following:

\begin{itemize}

\item{} The dwarf ellipticals in Fornax are found to have [Fe/H] values
ranging from $-$1.6 to $-$0.4 with a mean of $-$1.0.  The ages of dwarfs
ellipticals have a mean of 10$\pm$1 Gyrs.  For comparison, the
color-magnitude relation for bright ellipticals ($M_B < -18$) converts
into a range of [Fe/H] from $-$0.5 at the faint end to $+$0.3 for the most
massive systems, in agreement with previous studies.  These values are
based on luminosity-weighted colors corrected by our multi-metallicity
models.  The $\Delta(bz-yz)$ index indicates that the bright ellipticals
in clusters have a mean age of 13 Gyrs, on average 3 Gyrs older than
dwarfs, with very little scatter (i.e. coeval, see Kuntschner 2000).

\item{} The [Mg/Fe] ratio reflects the abundance of $\alpha$ elements to
Fe which varys due to the relative contribution from Type Ia and Type II
SN.  The data for dwarf ellipticals indicates [Mg/Fe] values slightly
higher than Galactic globular cluster, but near solar.  The $vz-yz$,
Mg$_2$ relationship for bright ellipticals indicates a trend of increasing
[$\alpha$/Fe] with mass.

\item{} The Fornax dwarf ellipticals do not, as a group, follow a
mass-metallicity relation found for bright ellipticals.  However, there
is good evidence to indicate that the dwarfs divide into high and low
metallicities sequences based on mean age.  The younger dwarfs have higher
metallicities per mass which suggests a longer, or more intensive, period of
star formation and resulting chemical evolution before the onset of
galactic winds.

\item{} There is a clear abundance and age gradient for Fornax dwarfs
within the cluster environment.  The gradients are such that older,
metal-poor dwarfs inhabit the central core of the cluster and the younger,
metal-richer dwarfs occupy the halo.  

\end{itemize}

As pointed out in Trager \etal (2000), there are numerous scenarios which
can lead to either an increasing [Fe/H] or an increasing [$\alpha$/Fe] with galaxy
mass.  However, it is difficult to resolve both effects with either a
simple increase in the initial star formation duration, or with earlier
and stronger galactic winds for low mass systems.  With respect to the
Fornax dwarfs, a somewhat clear picture presents itself due to the fact
that the older dwarfs have higher metallicities and the [$\alpha$/Fe]
values for dwarfs in general are near solar.  This suggests that star
formation in dwarf ellipticals occurs in a long, weak initial burst,
sufficiently long to acquire low [$\alpha$/Fe] values from Type Ia SN
enrichment, but also sufficiently low in strength to inhibit the onset of a
galactic wind.  Any extrapolation of this style of star formation to
bright ellipticals requires that the duration of initial star formation
decrease with increasing galaxy mass, which reconciles the increasing
[$\alpha$/Fe] ratio with mass.  But, the strength of the burst must also
increase with galaxy mass to increase the number of stellar generations in
order to reconcile the increase in [Fe/H] with galaxy mass.  A stronger
burst will inject more energy at a higher rate and, thus, produce an
earlier galactic wind.

The remaining mystery for the chemical evolution of dwarf ellipticals is
the extremely large spread in [Fe/H] values per mass bin.  If star
formation in dwarfs is halted by a galactic wind, then the mass
metallicity relation should continue into the dwarf regime.  However, the
galactic wind idea for dwarfs has come into question by Mac Low \& Ferrara
(1999) who propose, based on analytic and numerical simulations, that
galactic winds in dwarfs are impossible for masses above $10^9 M_{\sun}$.
Their model is primarily motivated by the concept that the hot gas within
the SN superbubble is easily heated to escape velocity, but the
surrounding cold gas must also be sweep up and accelerated, a much more
difficult energy requirement.  Without a galactic wind, star formation
duration would seem to be the controlling factor in determining the final
metallicity of a dwarf galaxy.  And this is confirmed in Figure 13 such
that the dwarfs with the longest epoch of star formation have the highest
metallicities.

Lastly, the information culled from Figure 14 suggests a strong
environmental influence on the star formation history of dwarf
ellipticals.  While it has been proposed that dwarf galaxies may
contribute some fraction of the ICM in rich clusters (Nath \& Chiba 1995,
Trentham 1994), there is no direct evidence that metal enriched gas has
been removed from cluster dwarf galaxies.   Without interpretation, the
correlation of metallicity and age indicates that star formation is
halted by some mechanism, prematurely, for core galaxies (old age) which,
in turn, ceases chemical evolution (lower metallicities).  The most likely
candidate for this mechanism is ram pressure stripping (Gunn \& Gott
1972); however, alternative scenarios, such as galaxy harassment (Moore
\etal 1996), have been proposed (see Skillman \& Bender 1995 for a
review).  Further observations into the spread of metallicity and age
within a cluster and addition knowledge of the range of [$\alpha$/Fe]
among cluster dwarfs will be required to better define the problem.

\acknowledgements

The authors wish to thank the directors and staff of CTIO, Steward and
Lowell Observatories for granting time for this project.  Financial
support from  Austrian Fonds zur Foerderung der Wissenschaftlichen
Forschung is gratefully acknowledged.  We also wish to thank the referee,
Scott Trager, for his numerous suggestions and comments that strengthened
the presentation of our data.

\begin{deluxetable}{lccccccccc}
\tablecolumns{10}
\small
\tablewidth{0pt}
\tablecaption{Galactic Globular Clusters}
\tablehead{
\colhead{Cluster} & 
\colhead{$uz-vz$} &
\colhead{$bz-yz$} &
\colhead{$vz-yz$} &
\colhead{$mz$} &
\colhead{Age} &
\colhead{[Fe/H]} &
\colhead{$E(B-V)$} &
\colhead{HB} &
\colhead{$\Delta(bz-yz)$} \nl
\colhead{} & 
\colhead{} & 
\colhead{} & 
\colhead{} & 
\colhead{} & 
\colhead{(Gyrs)} & 
\colhead{} & 
\colhead{} & 
\colhead{ratio} &
\colhead{} \nl
}
\startdata
~104&$+$0.64&$+$0.28&$+$0.51&$-$0.05&  9.2$\pm$1.0 & $-$0.70 & 0.04 & $-$0.99 & +0.067  \nl
~288&$+$0.51&$+$0.31&$+$0.31&$-$0.31&  8.8$\pm$0.8 & $-$1.07 & 0.03 & $+$0.98 & \nodata \nl
~362&$+$0.58&$+$0.27&$+$0.37&$-$0.17&  8.7$\pm$0.8 & $-$1.15 & 0.04 & $-$0.87 & +0.023  \nl
1851&$+$0.62&$+$0.26&$+$0.35&$-$0.17&  7.9$\pm$0.8 & $-$1.22 & 0.02 & $-$0.36 & +0.052  \nl
1904&$+$0.60&$+$0.20&$+$0.22&$-$0.20& 10.1$\pm$1.1 & $-$1.37 & 0.01 & $+$0.89 & +0.037  \nl
3201&$+$0.62&$+$0.27&$+$0.31&$-$0.23&  9.9$\pm$0.8 & $-$1.23 & 0.21 & $+$0.08 & -0.010  \nl
4147&$+$0.58&$+$0.16&$+$0.13&$-$0.19& 10.5$\pm$0.5 & $-$1.83 & 0.02 & $+$0.55 & +0.002  \nl
4372&$+$0.40&$+$0.14&$+$0.07&$-$0.21& 11.5$\pm$1.0 & $-$2.09 & 0.50 & $+$1.00 & -0.020  \nl
4590&$+$0.53&$+$0.17&$+$0.07&$-$0.27& 11.4$\pm$1.0 & $-$1.99 & 0.05 & $+$0.17 & -0.034  \nl
4833&$+$0.54&$+$0.18&$+$0.14&$-$0.22&    \nodata   & $-$1.58 & 0.33 & $+$0.93 & \nodata \nl
5024&$+$0.50&$+$0.14&$+$0.13&$-$0.15& 11.4$\pm$0.5 & $-$1.99 & 0.02 & $+$0.81 & -0.004  \nl
5053&$+$0.31&$-$0.04&$-$0.01&$+$0.07&    \nodata   & $-$2.29 & 0.04 & $+$0.52 & \nodata \nl
5139&$+$0.53&$+$0.22&$+$0.24&$-$0.20& 10.5$\pm$0.5 & $-$1.62 & 0.12 & $+$0.75 & -0.023  \nl
5272&$+$0.54&$+$0.21&$+$0.28&$-$0.14& 10.1$\pm$1.1 & $-$1.34 & 0.01 & $+$0.08 & +0.032  \nl
5286&$+$0.56&$+$0.20&$+$0.19&$-$0.21& 11.0$\pm$0.5 & $-$1.67 & 0.24 & $+$0.80 & -0.012  \nl
5466&$+$0.52&$+$0.07&$-$0.01&$-$0.15&    \nodata   & $-$2.22 & 0.00 & $+$0.58 & \nodata \nl
5634&$+$0.49&$+$0.13&$+$0.19&$-$0.07&    \nodata   & $-$1.82 & 0.05 & $+$0.90 & \nodata \nl
5897&$+$0.58&$+$0.22&$+$0.20&$-$0.24& 10.1$\pm$1.1 & $-$1.59 & 0.09 & $+$0.86 & -0.019  \nl
5904&$+$0.52&$+$0.18&$+$0.24&$-$0.12&  9.9$\pm$0.7 & $-$1.11 & 0.03 & $+$0.31 & +0.100  \nl
5986&$+$0.58&$+$0.22&$+$0.19&$-$0.23&    \nodata   & $-$1.58 & 0.27 & $+$0.97 & \nodata \nl
6101&$+$0.57&$+$0.23&$+$0.17&$-$0.25& 10.9$\pm$1.1 & $-$1.82 & 0.05 & $+$0.84 & -0.066  \nl
6171&$+$0.32&$+$0.20&$+$0.51&$+$0.11& 10.4$\pm$1.0 & $-$1.04 & 0.33 & $-$0.73 & \nodata \nl
6205&$+$0.50&$+$0.20&$+$0.25&$-$0.17& 10.3$\pm$0.9 & $-$1.39 & 0.02 & $+$0.97 & +0.034  \nl
6218&$+$0.44&$+$0.17&$+$0.28&$-$0.06& 10.5$\pm$1.0 & $-$1.48 & 0.19 & $+$0.97 & +0.049  \nl
6229&$+$0.52&$+$0.19&$+$0.31&$-$0.07&  9.5$\pm$0.5 & $-$1.43 & 0.00 & $+$0.24 & +0.038  \nl
6254&$+$0.32&$+$0.18&$+$0.30&$-$0.06& 10.1$\pm$1.1 & $-$1.41 & 0.28 & $+$0.98 & +0.051  \nl
6352&$+$0.67&$+$0.30&$+$0.53&$-$0.07&  9.4$\pm$0.6 & $-$0.64 & 0.21 & $-$1.00 & +0.057  \nl
6397&$+$0.54&$+$0.16&$+$0.07&$-$0.25& 11.4$\pm$1.1 & $-$1.82 & 0.18 & $+$0.98 & +0.004  \nl
6584&$+$0.56&$+$0.21&$+$0.22&$-$0.20& 10.1$\pm$1.0 & $-$1.49 & 0.10 & $-$0.15 & +0.008  \nl
6652&$+$0.66&$+$0.29&$+$0.50&$-$0.08&  8.0$\pm$1.1 & $-$0.96 & 0.05 & $-$1.00 & +0.034  \nl
6656&$+$0.48&$+$0.27&$+$0.31&$-$0.23&    \nodata   & $-$1.48 & 0.34 & $+$0.91 & \nodata \nl
6681&$+$0.52&$+$0.25&$+$0.25&$-$0.25&    \nodata   & $-$1.51 & 0.07 & $+$0.96 & \nodata \nl
6715&$+$0.50&$+$0.25&$+$0.33&$-$0.17&    \nodata   & $-$1.59 & 0.14 & $+$0.87 & \nodata \nl
6723&$+$0.56&$+$0.26&$+$0.36&$-$0.18&  9.7$\pm$0.5 & $-$1.12 & 0.05 & $-$0.08 & +0.018  \nl
6752&$+$0.54&$+$0.21&$+$0.17&$-$0.25&  9.6$\pm$1.1 & $-$1.42 & 0.04 & $+$1.00 & +0.019  \nl
6760&$-$0.12&$+$0.32&$+$0.70&$+$0.06&  9.2$\pm$0.5 & $-$0.52 & 0.77 & $-$1.00 & +0.056  \nl
6809&$+$0.72&$+$0.21&$+$0.20&$-$0.22& 12.0$\pm$1.0 & $-$1.81 & 0.07 & $+$0.87 & -0.044  \nl
6864&$+$0.62&$+$0.25&$+$0.35&$-$0.15&    \nodata   & $-$1.32 & 0.16 & $-$0.42 & \nodata \nl
6981&$+$0.62&$+$0.25&$+$0.28&$-$0.22&    \nodata   & $-$1.40 & 0.03 & $+$0.14 & \nodata \nl
7089&$+$0.63&$+$0.22&$+$0.24&$-$0.20&    \nodata   & $-$1.62 & 0.02 & $+$0.96 & \nodata \nl
7099&$+$0.58&$+$0.15&$+$0.07&$-$0.23& 11.5$\pm$1.0 & $-$1.91 & 0.03 & $+$0.89 & -0.001  \nl
\enddata
\tablecomments{
Col. (1): Cluster NGC number. 
Col. (2): extinction corrected $uz-vz$ color. 
Col. (3): extinction corrected $bz-yz$ color.
Col. (4): extinction corrected $vz-yz$ color. 
Col. (5): extinction corrected $mz$ index. 
Col. (6): cluster age in Gyrs (Salaris \& Weiss 1998).
Col. (7): mean cluster [Fe/H] from Carretta \& Gratton 1997 or Harris (1996).
Col. (8): $E(B-V)$ from Harris (1996). 
Col. (9): horizontal branch index from Lee, Demarque \& Zinn (1994).
Col. (10): the age index, $\Delta(bz-yz)$ (see text).
}
\end{deluxetable}

\begin{deluxetable}{lrrrr}
\tablecolumns{5}
\small
\tablewidth{0pt}
\tablecaption{Multi-Metallicity Models}
\tablehead{
\colhead{$f_m=$} & 
\colhead{1.0} &
\colhead{0.9} &
\colhead{0.8} &
\colhead{0.7} \nl
\colhead{} & 
\colhead{(\%)} &
\colhead{(\%)} &
\colhead{(\%)} &
\colhead{(\%)} \nl
}
\startdata
[Fe/]=$-$2.3 &  0.8 &  1.0 &  1.4 &  1.9 \nl
[Fe/]=$-$1.7 &  2.8 &  3.9 &  6.3 & 12.6 \nl
[Fe/]=$-$0.7 &  7.8 & 14.8 & 29.7 & 60.5 \nl
[Fe/]=$-$0.4 & 11.8 & 21.0 & 37.6 & 25.0 \nl
[Fe/]=~~0.0    & 24.5 & 39.9 & 25.0 &  0.0 \nl
[Fe/]=$+$0.4 & 52.3 & 19.4 &  0.0 &  0.0 \nl
$<$[Fe/H]$>$ & $+$0.21 & $+$0.00 & $-$0.34 & $-$0.65 \nl
[Fe/H]$_{vz-yz}$ & $-$0.10 & $-$0.37 & $-$0.65 & $-$0.91 \nl
\enddata
\end{deluxetable}

\begin{deluxetable}{lccccccc}
\tablecolumns{8}
\small
\tablewidth{0pt}
\tablecaption{Fornax Dwarfs}
\tablehead{
\colhead{FCC} & 
\colhead{$uz-vz$} &
\colhead{$bz-yz$} &
\colhead{$vz-yz$} &
\colhead{$mz$} & 
\colhead{[Fe/H]} & 
\colhead{$B_T$} &
\colhead{$\Delta(bz-yz)$} \nl
}
\startdata
 85 & 0.31 & 0.39 & 0.70 & 1.01 & $-$0.08 & 16.3 & $+$0.027 \nl
116 & 0.44 & 0.37 & 0.65 & 1.09 & $-$0.09 & 16.9 & $+$0.024 \nl
133 & 0.36 & 0.38 & 0.49 & 0.85 & $-$0.27 & 14.8 & $-$0.058 \nl
136 & 0.24 & 0.35 & 0.64 & 0.88 & $-$0.06 & 16.1 & $+$0.039 \nl
137 & 0.17 & 0.43 & 0.58 & 0.75 & $-$0.28 & 17.5 & $-$0.068 \nl
150 & 0.62 & 0.25 & 0.55 & 1.17 & $+$0.05 & 15.7 & $+$0.098 \nl
157 & 0.42 & 0.23 & 0.37 & 0.79 & $-$0.09 & 18.3 & $+$0.037 \nl
160 & 0.67 & 0.35 & 0.53 & 1.20 & $-$0.17 & 17.7 & $-$0.011 \nl
178 & 0.72 & 0.28 & 0.46 & 1.18 & $-$0.10 & 17.2 & $+$0.027 \nl
181 & 1.14 & 0.28 & 0.22 & 1.36 & $-$0.34 & 17.2 & $-$0.080 \nl
188 & 0.74 & 0.34 & 0.47 & 1.21 & $-$0.21 & 16.1 & $-$0.027 \nl
189 & 0.30 & 0.22 & 0.42 & 0.72 & $-$0.02 & 18.8 & $+$0.070 \nl
195 & 0.77 & 0.27 & 0.39 & 1.16 & $-$0.15 & 16.7 & $+$0.006 \nl
202 & 0.68 & 0.30 & 0.50 & 1.18 & $-$0.10 & 15.3 & $+$0.025 \nl
203 & 0.67 & 0.29 & 0.46 & 1.13 & $-$0.12 & 15.5 & $+$0.017 \nl
207 & 0.60 & 0.24 & 0.35 & 0.95 & $-$0.13 & 15.9 & $+$0.018 \nl
211 & 0.59 & 0.28 & 0.35 & 0.94 & $-$0.21 & 16.3 & $-$0.022 \nl
222 & 0.71 & 0.33 & 0.51 & 1.22 & $-$0.15 & 15.6 & $+$0.000 \nl
223 & 0.94 & 0.26 & 0.36 & 1.30 & $-$0.16 & 16.2 & $+$0.003 \nl
230 & 0.72 & 0.25 & 0.39 & 1.11 & $-$0.11 & 17.2 & $+$0.026 \nl
231 & 0.68 & 0.30 & 0.52 & 1.20 & $-$0.08 & 18.4 & $+$0.036 \nl
245 & 0.61 & 0.28 & 0.44 & 1.05 & $-$0.12 & 16.0 & $+$0.019 \nl
266 & 0.66 & 0.34 & 0.53 & 1.19 & $-$0.15 & 15.9 & $-$0.001 \nl
274 & 0.68 & 0.25 & 0.37 & 1.05 & $-$0.13 & 16.5 & $+$0.017 \nl
293 & 0.90 & 0.29 & 0.46 & 1.36 & $-$0.12 & 17.6 & $+$0.017 \nl
296 & 0.66 & 0.31 & 0.43 & 1.09 & $-$0.19 & 16.3 & $-$0.015 \nl
298 & 0.60 & 0.28 & 0.39 & 0.99 & $-$0.17 & 16.6 & $-$0.004 \nl
\enddata
\tablecomments{
Col. (1): FCC number.
Col. (2): extinction corrected $uz-vz$ color.
Col. (3): extinction corrected $bz-yz$ color.
Col. (4): extinction corrected $vz-yz$ color.
Col. (5): extinction corrected $mz$ color.
Col. (6): color determined [Fe/H], corrected to our multi-metallicity models.
Col. (7): apparent blue magnitude taken from Ferguson (1989).
Col. (8): the age index, $\Delta(bz-yz)$.
}
\end{deluxetable}

%
%
%
%
%
%
%
%
%
%
%
%
%

\end{document}